%% file: main.tex
\documentclass[sigconf]{acmart}

\AtBeginDocument{%
  }

\copyrightyear{2026}
\acmYear{2026}
\setcopyright{cc}
\setcctype{by-nc-nd}
\acmConference[CHI '26]{Proceedings of the 2026 CHI Conference on Human Factors in Computing Systems}{April 13--17, 2026}{Barcelona, Spain}
\acmBooktitle{Proceedings of the 2026 CHI Conference on Human Factors in Computing Systems (CHI '26), April 13--17, 2026, Barcelona, Spain}
\acmPrice{}
\acmDOI{10.1145/3772318.3791953}
\acmISBN{979-8-4007-2278-3/2026/04}

\usepackage{tabularx}
\usepackage{array}
\usepackage{multirow} 
\usepackage{xcolor}
\definecolor{nodeblue}{RGB}{70,130,180}        
\definecolor{conceptorange}{RGB}{255,140,0}    
\definecolor{edgegreen}{RGB}{34,139,34}        
\definecolor{processbrown}{RGB}{160,82,45}

\definecolor{lightgray}{RGB}{239,239,239}
\definecolor{mediumgray}{RGB}{204,204,204}
\definecolor{darkred}{HTML}{7e0f12}
\definecolor{darkgreen}{rgb}{0.0, 0.5, 0.0}
\definecolor{purple}{HTML}{7262ac}
\definecolor{softpink}{HTML}{FFD4DB}
\definecolor{softseafoam}{HTML}{ABD4D1}

\begin{document}

\title{Designing Computational Tools for Exploring Causal Relationships in Qualitative Data}




\author{Han Meng}
\email{han.meng@u.nus.edu}
\orcid{0009-0003-2318-3639}
\affiliation{
  \institution{Department of Computer Science, National University of Singapore}
  \streetaddress{21 Lower Kent Ridge Road}
  \country{Singapore}
  \postcode{119077}
}

\author{Qiuyuan Lyu}
\email{qiuyuan@u.nus.edu}
\orcid{0009-0009-2472-6272}
\affiliation{
  \institution{Department of Computer Science, National University of Singapore}
  \streetaddress{21 Lower Kent Ridge Road}
  \country{Singapore}
  \postcode{119077}
}

\author{Peinuan Qin}
\email{e1322754@u.nus.edu}
\orcid{0000-0002-8737-8369}
\affiliation{
  \institution{Department of Computer Science, National University of Singapore}
  \streetaddress{21 Lower Kent Ridge Road}
  \country{Singapore}
  \postcode{119077}
}

\author{Yitian Yang}
\email{yang.yitian@u.nus.edu}
\orcid{0009-0000-7530-2116}
\affiliation{
  \institution{Department of Computer Science, National University of Singapore}
  \streetaddress{21 Lower Kent Ridge Road}
  \country{Singapore}
  \postcode{119077}
}

\author{Renwen Zhang}
\email{renwen.zhang@ntu.edu.sg}
\orcid{0000-0002-7636-9598}
\affiliation{
  \institution{Wee Kim Wee School of Communication and Information, Nanyang Technological University}
  \streetaddress{50 Nanyang Avenue}
  \country{Singapore}
  \postcode{639798}
}

\author{Wen-Chieh Lin}
\email{wclin@cs.nctu.edu.tw}
\orcid{0000-0002-9704-5373}
\affiliation{
  \institution{Dept of Computer Science, National Yang Ming Chiao Tung University}
  \city{Hsinchu}
  \country{Taiwan}
}

\author{Yi-Chieh Lee}
\email{yclee@nus.edu.sg}
\orcid{0000-0002-5484-6066}
\affiliation{
  \institution{Department of Computer Science, National University of Singapore}
  \streetaddress{21 Lower Kent Ridge Road}
  \country{Singapore}
  \postcode{119077}
}





\renewcommand{\shortauthors}{}


\begin{abstract}
\begin{figure*}[th]
    \centering
    \includegraphics[width=1\linewidth]{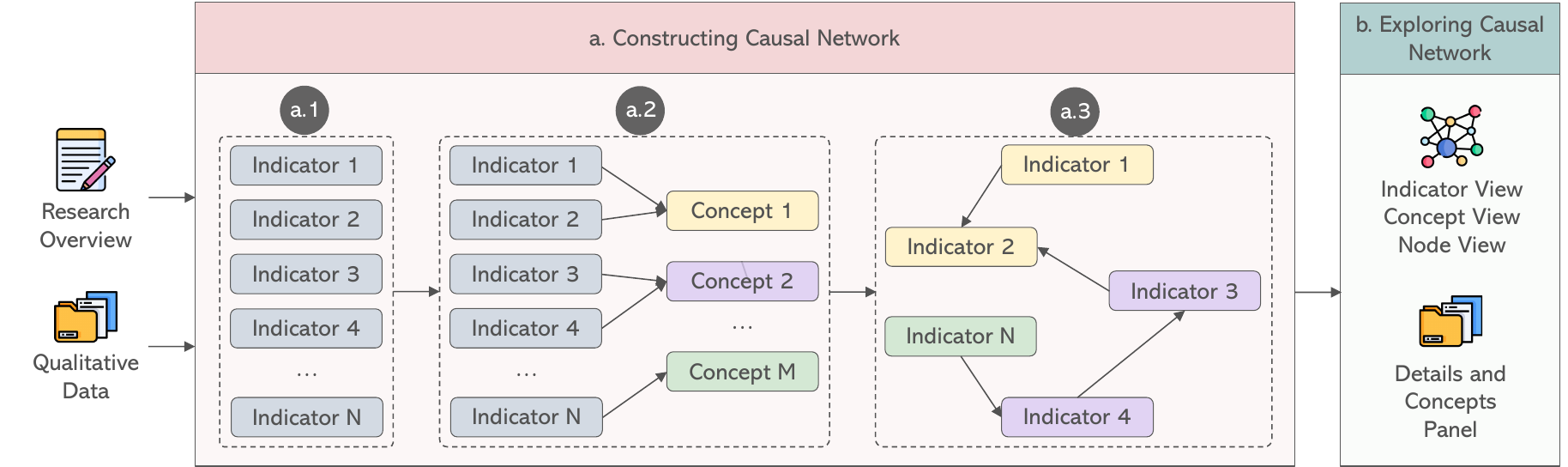}
    \caption{An overview of \texttt{QualCausal}. \texttt{QualCausal} consists of (a) \textbf{Causal Network Construction} through three steps: (a.1) extracting indicators from qualitative data based on the user-provided research overview, (a.2) interactively creating and mapping concepts with user control, and (a.3) extracting causal relationships; and (b) \textbf{Causal Network Exploration} through multiple coordinated views (i.e., \textit{Indicator View}, \textit{Concept View}, and \textit{Node View}), with \textit{Details and Concepts Panels} providing access to source data.}
    \Description{Flowchart diagram showing the QualCausal workflow organized into two main phases. The left side shows 'Causal Network Construction' with three sequential steps labeled a.1, a.2, and a.3. Step a.1, 'Indicator Extraction', displays a list of indicators (Indicator 1, 2, 3, 4, ..., N) extracted from qualitative data based on the user-provided research overview. Step a.2 'Concept Creation and Mapping' shows these indicators being mapped to concepts (Concept 1, 2, ..., M) with arrows connecting indicators to their assigned concepts. Step a.3 'Causal Relationship Extraction' presents a network diagram where indicators are connected by directional arrows representing causal relationships. The right side shows 'Causal Network Exploration' featuring multiple coordinated views: an Indicator View displaying the full causal network, a Concept View showing abstracted concept-level causal relationships, a Node View presenting localized subnetworks, and a Details and Concepts Panel for accessing source data.}
    \label{fig:teaser}
\end{figure*}

  Exploring causal relationships for qualitative data analysis in HCI and social science research enables the understanding of user needs and theory building. 
  However, current computational tools primarily characterize and categorize qualitative data; the few systems that analyze causal relationships either inadequately consider context, lack credibility, or produce overly complex outputs. 
  We first conducted a formative study with 15 participants interested in using computational tools for exploring causal relationships in qualitative data to understand their needs and derive design guidelines. 
  Based on these findings, we designed and implemented \texttt{QualCausal}, a system that extracts and illustrates causal relationships through interactive causal network construction and multi-view visualization. 
  A feedback study ($n=15$) revealed that participants valued our system for reducing the analytical burden and providing cognitive scaffolding, yet navigated how such systems fit within their established research paradigms, practices, and habits. 
  We discuss broader implications for designing computational tools that support qualitative data analysis.
\end{abstract}




\maketitle

\input{sections/1_intro}

\input{sections/2_related_work}

\input{sections/3_formative_study}

\input{sections/4_system_design}

\input{sections/5_feedback_study}

\input{sections/6_discussion}

\begin{acks}

Our research has been supported by the Ministry of Education, Singapore (A-8002610), NUS Centre for Computational Social Science and Humanities (A-8002954), and Nanyang Technological University Start-up Grant (NAP\_SUG 025564-00001). 
We are grateful to all the reviewers for their valuable comments and suggestions that helped improve this paper, and to all the participants whose time and efforts made this research possible.
\end{acks}

\bibliographystyle{ACM-Reference-Format}
\bibliography{sample-base}



\end{document}

%% file: sections/1_intro.tex
\section{Introduction}
\label{sec:intro}

Qualitative data analysis plays a crucial role in HCI and social science research for understanding people, facilitating insights into social processes, and supporting theory development \cite{hci_qual_lazar_2017}.
Common analytical approaches to achieving these goals unfold by answering ``\textit{What do the data look like?}'' and ``\textit{Why/What do the data mean?}'' \cite{causal_qual_katz_2001, causal_qual_katz_2002}. 
Exploring \textbf{causal relationships} in qualitative data is therefore important, as it can help predict human behaviors and untangle complex social phenomena \cite{grounded_theory_larossa_2005, causal_maxwell_2004}. 
Particularly, the rich contextual details, processes, and meanings embedded in qualitative data \cite{grounded_theory_maxwell_2007} can inform the generation of exploratory hypotheses for quantitative causal identification methods like randomized experiments \cite{causal_shadish_2002} and structural models \cite{causality_pearl_2009, causal_spirtes_2000} to test.
Yet, the sheer volume and unstructured nature of qualitative data make analysis by hand demanding and labor-intensive \cite{ai_characterize_viz_huang_2025}.

In response, a growing collection of methods and systems has emerged to help users who would like to use computational tools understand \cite{computer_deductive_nelson_2021}, characterize \cite{tradition_computer_inductive_lennon_2021}, and categorize \cite{llm_inductive_lee_2024} large amounts of qualitative data. 
Topic modeling \cite{topic_visual_cui_2011} and text clustering \cite{traditional_nlp_inductive_parfenova_2024}, for example, uncover and group latent, similar content within the dataset.
Systems that integrate these approaches generate more comprehensible and interactive results by displaying topic models \cite{topic_visual_dou_2013}, quantifying words and phrases \cite{word_visual_rohrdantz_2012}, and highlighting named entities \cite{ner_visual_koch_2014}. 
More recent AI- and LLM-based systems help openly code extracted data features into abstract concepts \cite{llm_inductive_lam_2024, llm_inductive_zenimoto_2024}, organize and summarize concepts \cite{chalet_meng_2024}, and group data into coherent themes \cite{cgt_alqazlan_2025}. 
However, these methods and systems remain overly \textit{descriptive}, focusing on categorization and likely overlooking the need to tease apart antecedents, underlying facets, and consequences \cite{descriptive_boyd_2021}.

Several computational methods and tools have been preliminarily used to \textit{explain} qualitative data by exploring \textit{causal relationships} \cite{visual_cooccurrence_andrienko_2020, causal_nlp_feuerriegel_2025}. 
Some systems use network analysis to connect concepts based on co-occurrence patterns \cite{visual_cooccurrence_schwab_2024, visual_cooccurrence_chandrasegaran_2017}. 
Drawing from natural language processing methods for causal inference, such as Bayesian inference \cite{causal_nlp_nazaruka_2019} or LLM reasoning \cite{causal_nlp_newberry_2024}, other systems embed these approaches to create connections between theoretical constructs \cite{causal_llm_jalali_2024, causal_llm_hosseinichimeh_2024} or source data segments \cite{causal_system_powell_2025, causal_system_powell_2024}.
However, systems that extract causal relationships only at the conceptual level make it challenging to systematically explore how the automatically identified causal relationships are supported by the data. 
Conversely, systems that operate directly on raw text generate vast networks that are difficult to interpret and navigate, offering limited insight into theory building.

Visualization approaches have been shown to help address the needs for tracing, validating, organizing, and making sense of causal relationships \cite{causal_viz_yen_2019, causal_viz_yen_2019_2}. 
Prior research has focused on visualizing causal relationships in numerical data based on statistical tests through network structures, such as directed acyclic graphs (DAGs) \cite{causal_viz_kapler_2021, causal_viz_wang_2015}, and layered node-link diagrams \cite{causal_viz_fan_2024, causal_viz_xie_2020} and/or embedding causal-discovery algorithms into interactive user interfaces \cite{word_cloud_wang_2017}.
These approaches provide valuable insights into designing visual displays, yet they are intended for modeling causal relationships in quantitative datasets.

This work aims to develop systems that support exploring causal relationships in qualitative data and generating hypotheses.
We conducted a three-phase study. 
In Phase one, we conducted a formative study with 15 participants to understand their current methods, challenges, and envisioned features. 
Based on the findings, in Phase two, we developed \texttt{QualCausal}, a system that combines \textit{causal network} visualization with a user-controlled workflow for categorizing and causally connecting qualitative data. 
In Phase three, we conducted a feedback study ($n=15$) on \texttt{QualCausal}, which also facilitated broader conversations about how computational tools could better support qualitative data analysis.
Our findings revealed a tension between users valuing the speed and cognitive support of these tools and concerns about their influence on established practices and paradigms.

Our paper makes several contributions to the HCI community.
First, it establishes \textbf{design principles} for computational tools that support exploring causal relationships and generating hypotheses from qualitative data using computational methods and interactive visualizations.
Second, it introduces \texttt{QualCausal}, a novel system for constructing causal networks from qualitative data and exploring them, balancing automation with user autonomy and visualizing broad patterns, details, and source data.
Third, it discusses the \textbf{implications of designing computational tools} for qualitative data analysis and broader computational social science, focusing on how these tools affect analytical experiences and epistemological assumptions.

%% file: sections/2_related_work.tex
\section{Related Work}
\label{sec:rw}

\subsection{Qualitative Data Analysis and Causal Relationships}

Qualitative data analysis serves as an important research method in HCI and social science for interpreting data from interviews, focus groups, observations, and more \cite{qda_flick_2014}. 
The goal is to transform unstructured datasets into organized insights that help understand how social problems unfold, are construed, and perpetuated \cite{hci_qual_lazar_2017}.

Grounded Theory \cite{grounded_theory_larossa_2005} is one of the most commonly used analytical approaches \cite{grounded_theory_bryant_2010}, which is purposefully explanatory \cite{grounded_theory_larossa_2005}, facilitates theory building \cite{grounded_theory_charmaz_2006}, and can serve as inductive legwork prior to quantitative studies, which helps select variables and pathways \cite{grounded_theory_bennett_2016}.
It contains two main processes: \textit{categorization} (open coding) and \textit{connection to theorize} (axial/selective coding) \cite{grounded_theory_maxwell_2007, grounded_theory_larossa_2005}. 
\textit{Categorization} involves systematically breaking down textual data into discrete components and developing conceptual labels \cite{grounded_theory_larossa_2014}. 
The \textit{connection} phase explores how elements relate through questions of when, where, why, and under what conditions \cite{grounded_theory_larossa_2005}. 

Among these connections, \textit{causal relationships} \cite{causal_qual_katz_2001, causal_qual_katz_2002} have significant potential for qualitative data analysis. 
Qualitative data offers unique advantages for causal discovery because its \textit{process-oriented} \cite{causal_maxwell_2004} nature allows for the examination of social processes as observable sequences of symbolically mediated interactions \cite{causal_charmaz_2000}. 
This allows them to trace the actual, in-situ connections between events and their complex interactions within specific \textit{contexts} \cite{qda_tracy_2024}. 
Additionally, the capacity of qualitative data to capture \textit{meaning} enables understanding of human beliefs, values, and mental processes that influence behavior and shape causal explanations \cite{causal_maxwell_2021}.


Traditional causal relationship exploration in qualitative data involves linking antecedents and consequents \cite{causal_maxwell_2016}, searching for causal connecting words \cite{causal_blalock_2018}, and examining how elements influence each other \cite{causal_education_maxwell_2004}. 
However, this processual explanation requires intensive, long-term involvement and narrative approaches, making manual methods demanding and time-consuming \cite{cld_rajah_2025, causal_pyrko_2018}.

\subsection{Computational Support for Qualitative Data Analysis}

\subsubsection{Computational Support for Characterizing and Categorizing Qualitative Data}

Computational methods and tools begin to support qualitative data analysis \cite{ai_qda_chen_2018, ai_qda_jiang_2021, ai_qda_muller_2016}. 
Previous studies have shown that unsupervised machine learning techniques like topic modeling and clustering algorithms (e.g., $k$-means \cite{computer_deductive_nelson_2021}, HDBSCAN \cite{traditional_nlp_inductive_parfenova_2024}) can help characterize large datasets efficiently \cite{tradition_computer_inductive_lennon_2021, ai_qc_eads_2021}. 
Integrating these methods into systems makes the results more accessible and interactive \cite{ai_inductive_bakharia_2016, text_visual_kim_2020}. 
These systems use visual analytics techniques, such as coordinated panels \cite{topic_visual_cui_2011}, color coding \cite{topic_visual_dou_2013}, and graphical elements \cite{text_visual_park_2017}, to display topic models \cite{topic_visual_chuang_2012, topic_visual_wei_2010, topic_visual_sun_2014}, word-frequency patterns \cite{word_visual_rohrdantz_2012}, event sequences \cite{event_visual_dou_2012}, named entities \cite{ner_visual_koch_2014}, and semantic weights \cite{word_cloud_cui_2010, word_cloud_wang_2017}. 
StructVizor \cite{ai_characterize_viz_huang_2025}, for example, automatically parses textual data through field alignment and clustering, presenting results via interactive tabular views and heatmaps.

The emergence of AI/LLMs provides further support for categorizing and conceptualizing qualitative data based on these extracted features \cite{inductive_paoli_2024}. 
Previous work has shown that LLMs can assist with most stages of this inductive process, including generating initial codes \cite{llm_inductive_lam_2024, llm_inductive_zenimoto_2024}, identifying concepts \cite{llm_inductive_lee_2024}, and reviewing and refining those concepts \cite{inductive_labelling_dai_2023}. 
Specifically, in the initial open-coding phase of Grounded Theory, LLMs excel at suggesting and generating descriptive codes for data segments \cite{llm_inductive_bryda_2024}. 
For higher-level concept development, AI/LLMs can help group related codes based on shared properties \cite{inductive_morgan_2023}, propose names for recurring patterns \cite{inductive_yan_2024}, flag potentially important data that may have been overlooked \cite{chalet_meng_2024}, and suggest similar data segments for comparison and validation \cite{cgt_alqazlan_2025}. 
These proposed methodologies have been integrated into various systems \cite{open_coding_visual_rietz_2021, open_coding_drouhard_2017, open_coding_ganji_2018}, such as PATAT \cite{patat_gebreegziabher_2023}, Scholastic \cite{scholastic_hong_2022}, CollabCoder \cite{open_coding_visual_gao_2024}, CoAICoder \cite{open_coding_visual_gao_2023}, and ScholarMate \cite{ScholarMate_ye_2025} that can expand code repertoires \cite{open_coding_visual_muthukrishnan_2019}, provide statistical breakdowns of code usage \cite{open_coding_visual_nguyen_2015}, and offer on-demand suggestions for grouping related codes into abstract concepts \cite{open_coding_visual_gao_2024}.

Yet, these methods and systems primarily support \textit{describing} qualitative data \cite{descriptive_boyd_2021} through activities like classifying and quantifying words and phrases, while decomposing the antecedents, triggers, and consequences of events runs the risk of being oversimplified or altogether overlooked.
Our work expands on these ideas by proposing a system that uses computational methods to help users \textit{explain} qualitative data.

\subsubsection{Computational Support for Causally Connecting and Theorizing Qualitative Data}

There has been an increasing turn to computational methods and tools to causally connect concepts and develop theoretical explanations from qualitative data \cite{causal_llm_jalali_2024, causal_llm_hosseinichimeh_2024, causal_system_husain_2021, causal_system_powell_2025, causal_system_powell_2024}. 
The most common approach to exploring causal relationships relies on \textit{co-occurrence patterns} \cite{visual_cooccurrence_andrienko_2020}.
When text segments instantiating two concepts appear in the same sentence, systems create undirected edges between them \cite{visual_cooccurrence_pokorny_2018}, or directed edges based on chronological ordering from earlier to later mentions \cite{visual_cooccurrence_schwab_2024}. 
\textit{Tools} such as recent versions of NVivo\footnote{\url{https://help-nv.qsrinternational.com/20/mac/Content/queries/queries.htm}} and \textit{network analysis} toolkits like Gephi, Cytoscape, and \textit{systems} built on them \cite{visual_cooccurrence_chandrasegaran_2017} follow this approach when applied to qualitative data.
However, relying solely on co-occurrence patterns can overlook semantic, temporal, and contextual nuances, potentially creating misleading causal relationships.

Recent advances in natural language processing for causal inference and hypothesis generation \cite{hypothesis_park_2024, causal_graph_tong_2024} have opened up new possibilities, such as linguistic, semantic, and syntactical analysis \cite{causal_nlp_egami_2022, causal_nlp_sharp_2019}, Bayesian inference \cite{causal_nlp_nazaruka_2019}, and methods developed to identify temporal relations \cite{causal_inference_feder_2022}. 
Initially, these computational approaches were considered unreliable and immature for qualitative data analysis \cite{causal_nlp_kenzie_2025}; however, ongoing developments have made it possible to apply these approaches to behavioral science \cite{causal_nlp_feuerriegel_2025}, psychology \cite{causal_text_meng_2025}, and system dynamics \cite{causal_nlp_newberry_2024} to detect causal relationships from qualitative data.
Early implementations relied on simple causal keyword matching, searching for causal connectives like ``\textit{because},'' ``\textit{therefore},'' and ``\textit{as a result}'' based on linguistic inventories \cite{causal_nlp_newberry_2024, causal_nlp_altenberg_1984}. 
More recent approaches include prompting LLMs to ``\textit{identify key variables}'' and ``\textit{establish causal links between variables}'' from qualitative data \cite{causal_llm_jalali_2024, causal_llm_hosseinichimeh_2024}.
Only a few \textit{systems} have been developed for this purpose \cite{causal_system_husain_2021}. 
One example is CausalMap, which uses graph structures to display causal relationships between raw data segments \cite{causal_system_powell_2025, causal_system_powell_2024}.

However, existing systems face limitations. 
First, systems that only infer concept-level causal relationships \cite{causal_llm_jalali_2024, causal_llm_hosseinichimeh_2024} present verification and tracking challenges. 
Without mechanisms to facilitate tracing between extracted causal relationships and their supporting evidence in the source data, it becomes challenging to validate the generated hypotheses.
Second, systems that work solely with raw text segments \cite{causal_system_powell_2025, causal_system_powell_2024, causal_system_husain_2021} produce overly complex outputs (i.e., an enormous, difficult-to-interpret mapping) and provide little summative information or theoretical insights.
Building on this prior work, we introduce a system that provides global causal relationship views, traceability back to source data, and concept-level networks to inform theoretical development.


\subsection{Visualizations of Causal Relationships}

The challenges of tracing, validating, organizing, and interpreting causal relationships from qualitative data can be partially addressed through visual analytics approaches. 
Visual representations can reduce cognitive load \cite{causal_viz_xiong_2019} and provide interactive mechanisms when processing relational information \cite{causal_viz_yen_2019, causal_viz_yen_2019_2}.

Existing research on visualizing causal relationships has explored various representation strategies. 
DAGs remain the most common approach for depicting causal structures \cite{causal_viz_kapler_2021}. 
For instance, cause-and-effect direction is depicted with arrows and tapered lines; causal strength is represented with hue and line width, which can be given numeric values; and uncertainty is delineated with granularity, brightness, and fuzziness \cite{causal_viz_bae_2017}, as employed in systems like The Visual Causality Analyst \cite{causal_viz_wang_2015}, Causalvis \cite{causal_viz_guo_2023}, and Outcome-Explorer \cite{causal_viz_2021}. 
Layered node-link diagrams \cite{causal_viz_fan_2024, causal_viz_xie_2020_2} organize causal variables by temporal or hierarchical ordering, with CausalFlow \cite{causal_viz_xie_2020} being a notable implementation. 
Flow-based visualizations \cite{causal_viz_jin_2020} use stream-like representations to portray causal pathways and their relative strengths, as seen in ReactionFlow \cite{causal_viz_dang_2015} and DOMINO \cite{causal_viz_wang_2022}. 
Multiple coordinated views combine and link different visual perspectives and dimensions, while systems like Compass \cite{causal_viz_deng_2021} and VAC2 \cite{causal_viz_zhu_2024} use this approach alongside directed hypergraph representations to capture complex causation.

Although most existing visualization systems focus on causal relationships derived from quantitative, numeric data, these systems still provide valuable methodological insights.
Inspired by these works, we propose a computational tool that helps causal discovery and generates hypotheses from qualitative data by extracting and categorizing concepts, connecting them to explore causal relationships, and interactively visualizing these relationships.


In the following, we define the terms integral to this paper:

\begin{itemize}
   \item \colorbox{nodeblue!20}{indicators}: words, phrases, sentences, or series thereof in the analyzed materials that are usually highlighted as notes and keywords during the open coding phase \cite{grounded_theory_larossa_2005}.

   \item \colorbox{conceptorange!20}{concepts}: labels or names associated with one or more indicators, serving as categorical markers to identify and classify specific types of phenomena or entities \cite{grounded_theory_larossa_2005}.
   
   \item \colorbox{edgegreen!20}{causal relationships}: directed connections representing potential cause-and-effect pathways \cite{causal_maxwell_2004} at both indicator-level and concept-level.
   
   \item \colorbox{processbrown!20}{causal networks}: graph structures visualizing causal relationships at different abstraction levels. Indicator-level networks display data segments and connections, analogous to causal graphs in quantitative research \cite{causal_viz_guo_2023, causal_viz_fan_2024}. Concept-level networks present theoretical constructs as conceptual models for theory building \cite{hci_qual_lazar_2017}. It is important to note that these networks differ from causal loop diagrams (CLDs), which are archetypes commonly used in system dynamics \cite{cld_rajah_2025} to design interventions and study system behavior rather than to develop theory.
\end{itemize}

%% file: sections/3_formative_study.tex
\section{Formative Study}
\label{sec:formative}

\begin{figure*}[t]
    \centering
    \includegraphics[width=1\linewidth]{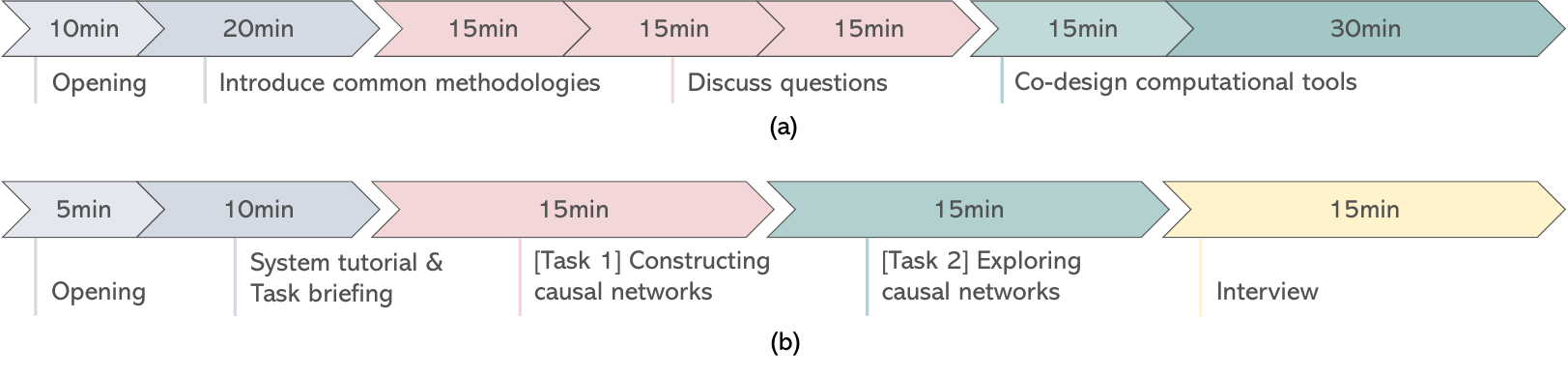}
    \caption{Processes of formative study and feedback study. (a) The formative study was a two-hour workshop consisting of an opening (10 minutes), introduction of common methods (20 minutes), group discussions on three questions (15 minutes each), and a co-design activity for computational tools (45 minutes). (b) The feedback study was a one-hour session consisting of an introduction and system tutorial (10 minutes), followed by two hands-on evaluation sessions (15 minutes each) with \texttt{QualCausal}, and concluding with an experience discussion (15 minutes).}
    \Description{Timeline diagram showing the procedures for two studies. Part (a) shows the formative study as a horizontal timeline with chevron-shaped segments indicating sequential phases over 2 hours total. The timeline begins with Opening (10 minutes), followed by Introduce common methodologies (20 minutes), then Discuss questions (15 minutes), another Discuss questions segment (15 minutes), a third Discuss questions segment (15 minutes), and concludes with Co-design computational tools (30 minutes). Part (b) shows the feedback study as a similar horizontal timeline over 1 hour total, consisting of Opening (5 minutes), System tutorial & Task briefing (10 minutes), Task 1 Constructing causal networks (15 minutes), Task 2 Exploring causal networks (15 minutes), and Interview (15 minutes). Both timelines use different colored segments to distinguish between different types of activities, with the chevron design indicating the sequential flow of each study phase.}
    \label{fig:study_design}
\end{figure*}

We conducted a formative study to identify the current challenges of exploring causal relationships with qualitative data, to understand how computational tools are being used, and to find ways in which these tools could provide better support. 
This study consisted of a need-finding workshop with 15 participants who had prior experience using computational tools for qualitative data analysis or who expressed interest in incorporating such tools into their practices.

\subsection{Participants}

We recruited participants through a \textbf{workshop} that brought together people from multiple countries who were interested in applying computational tools to analyze qualitative data.
Four university faculty members, two industrial researchers, and nine Ph.D. students participated in the workshop (Table \ref{tab:participants}). 
The participants came from diverse disciplines, including political science, sociology, communication studies, HCI, and more. 
Of the 15 participants, 11 were male, and 4 were female. 
Their ages ranged from 25 to 44, and their experience with qualitative data analysis ranged from less than one year to over ten years.

\begin{table*}
  \centering
  \small
  \caption{Participant demographics and study participation. F\# denotes faculty members, I\# denotes industrial researchers, and P\# denotes Ph.D. students.}
  \Description{Table showing participant demographics and study participation across two studies. Participants are categorized as faculty members (F1-F6), industrial researchers (I1-I5), and Ph.D. students (P1-P12), with gender indicated in parentheses. Age ranges span from 18-24 to 35-44 years old, with most participants in the 25-34 age range. Research fields include diverse disciplines such as Political Science, Computer Science, Sociology, Communication Studies, Psychology, HCI, Education, and Social Work. Qualitative data analysis experience varies from less than 1 year to more than 10 years, with most having 1-5 years of experience. The table tracks participation in two studies: 15 participants took part in the formative study (indicated by checkmarks), and 15 participants participated in the feedback study. For feedback study participants, additional notation indicates whether they had prior experience using computational tools for qualitative data analysis, with 'yes' indicating experience and 'no' indicating no prior experience. Some participants participated in both studies, while others participated in only one study.}
    \begin{tabularx}{\textwidth}{>{\centering\arraybackslash}p{0.10\textwidth} >{\centering\arraybackslash}p{0.15\textwidth} >{\raggedright\arraybackslash}p{0.24\textwidth} >{\centering\arraybackslash}p{0.20\textwidth} >{\centering\arraybackslash}p{0.08\textwidth} >{\centering\arraybackslash}p{0.10\textwidth}}
    \toprule
    \multirow{2}[1]{*}{\textbf{Participant ID}} & 
    \multirow{2}[1]{*}{\textbf{Age}} & 
    \multirow{2}[1]{*}{\textbf{Research Field}} & 
    \multirow{2}[1]{*}{\begin{minipage}[c]{0.18\textwidth}\centering\textbf{Qualitative Data Analysis Experience}\end{minipage}} & 
    \multicolumn{2}{c}{\textbf{Study Participation}} \\
    \cmidrule(lr){5-6}
    & & & & 
    \textbf{Formative} & 
    \textbf{Feedback} \\
    \midrule
    F1 (M) & 25-34 years old & Political Science, Computer Science, Public Policy & 1-2 years & \checkmark & \checkmark/yes \\
    F2 (M) & 35-44 years old & Sociology, Political Science, Communication Studies & 1-2 years & \checkmark & \checkmark/yes \\
    F3 (M) & 25-34 years old & Sociology, Demography & 1-2 years & \checkmark & \checkmark/no \\
    F4 (M) & 35-44 years old & Information Science/HCI & 6-10 years &  & \checkmark/yes \\
    F5 (F) & 35-44 years old & Social Work & More than 10 years &  & \checkmark/no \\
    F6 (M) & 35-44 years old & Linguistics & 3-5 years & \checkmark &  \\
    I1 (F) & 35-44 years old & Education & Less than 1 year & \checkmark &  \\
    I2 (F) & 35-44 years old & Psychology, Education & 3-5 years &  & \checkmark/no \\
    I3 (F) & 25-34 years old & Computer Science & 6-10 years & \checkmark & \checkmark/no \\
    I4 (F) & 35-44 years old & Communication Studies, Public Health & 6-10 years &  & \checkmark/no \\
    I5 (F) & 25-34 years old & Information Science/HCI & 6-10 years &  & \checkmark/yes \\
    P1 (F) & 18-24 years old & Communication Studies & 6-10 years &  & \checkmark/no \\
    P2 (M) & 25-34 years old & Business/Management, Economics & Less than 1 year & \checkmark &  \\
    P3 (M) & 25-34 years old & Communication Studies & Less than 1 year & \checkmark &  \\
    P4 (F) & 25-34 years old & Sociology, Communication Studies & 6-10 years &  & \checkmark/no \\
    P5 (M) & 25-34 years old & Psychology & 3-5 years & \checkmark &  \\
    P6 (M) & 25-34 years old & Political Science, Public Administration & 1-2 years & \checkmark &  \\
    P7 (F) & 25-34 years old & Information Science/HCI & 1-2 years & \checkmark &  \\
    P8 (F) & 25-34 years old & Communication Studies & 3-5 years & \checkmark & \checkmark/yes \\
    P9 (M) & 25-34 years old & Communication Studies & 3-5 years & \checkmark & \checkmark/yes \\
    P10 (M) & 25-34 years old & Sociology & 1-2 years & \checkmark & \checkmark/yes \\
    P11 (M) & 25-34 years old & Communication Studies & Less than 1 year & \checkmark &  \\
    P12 (F) & 25-34 years old & Communication Studies & 3-5 years &  & \checkmark/no \\
    \bottomrule
    \end{tabularx}%
    \vspace{0.5em}
    \raggedright
    \small
    \textit{Note:} In the \textit{Feedback} column, ``yes/no'' indicates whether participants have previous experience using computational tools for qualitative data analysis.

  \label{tab:participants}%
\end{table*}

\subsection{Procedure}
Our formative study workshop was two hours in-person, facilitated by three members of the research team. 
We began with a 10-minute session where participants introduced themselves and shared their professional backgrounds and experience with qualitative data analysis and computational methods. 
We then provided a 20-minute introduction to key approaches, covering widely used qualitative data analysis methods like Grounded Theory \cite{grounded_theory_larossa_2005} and common computational approaches such as topic modeling \cite{topic_modeling_rodriguez_2020} and NLP-based causal inference \cite{causal_text_meng_2025, causal_nlp_feuerriegel_2025}.

After the opening, we divided participants into four mixed groups of three to four people each, intentionally combining participants from different professional backgrounds. 
We posed three \textbf{discussion questions}: 1) their current methods for exploring causal relationships and generating hypotheses from qualitative data, 2) the challenges they encounter in this process, and 3) their vision for how computational tools could better support causal discovery in qualitative data analysis.
Each question received five minutes of group discussion time, followed by two and a half minutes for each group to share insights with all participants using post-it notes on the whiteboard.

In the final activity, participants individually sketched their envisioned system for qualitative data analysis using colored pens and paper. 
We chose \textbf{paper prototyping} to encourage rapid ideation without technical constraints \cite{paper_prototyping_2003}.
The exercise took 15 minutes, followed by two minutes per person (30 minutes total) to share and discuss their designs. 
We concluded by providing key takeaways and contact information for those interested in signing up for follow-up studies (i.e., feedback study). 
Figure \ref{fig:study_design} (a) illustrates the workshop procedure.

We video-recorded the workshop and transcribed it. 
We performed thematic analysis \cite{thematic_analysis_braun_2006} on the transcripts. 
The first author open-coded and affinity diagrammed the transcripts. 
The full research team then discussed and identified themes. 
We also conducted content analysis \cite{sketch_analysis_buxton_2007} on the participants' design sketches.

\begin{figure*}
    \centering
    \includegraphics[width=0.8\linewidth]{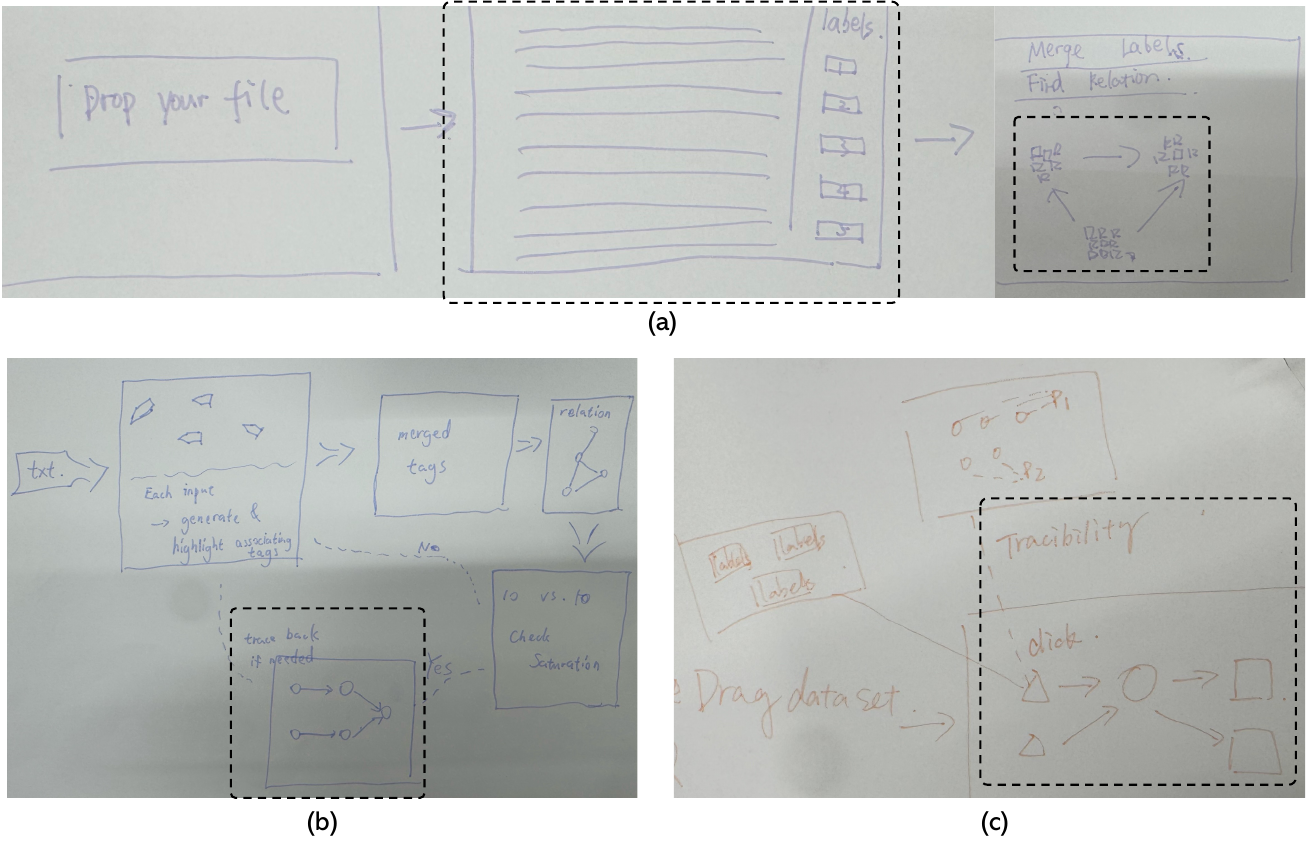}
    \caption{Paper prototypes created by participants during the formative study co-design activity. (a) A system that processes uploaded files through automated labeling and displays results as directed graphs. (b) A workflow where input text files are chunked into segments, tags are generated and highlighted for each segment, and the results are then organized into network structures that allow users to trace back to source data. (c) Drag-and-drop functionality with different node shapes representing different types of labels, where clicking nodes enables traceability to source data.}
    \Description{Three hand-drawn paper prototypes created by participants during the formative study co-design activity. Prototype (a) shows a linear workflow starting with "Drop your file" on the left, followed by a central panel with lined sections for text processing and small icons on the right side, leading to a final output panel on the far right showing interconnected nodes and labels with arrows indicating relationships. Prototype (b) displays a more complex multi-panel design with various rectangular sections containing handwritten text and diagrams, including workflow arrows, text segments, and network visualization components, with a highlighted section showing connected nodes and relationships. Prototype (c) presents a drag-and-drop interface concept with "drag data set" written prominently, featuring different geometric shapes (circles, rectangles, triangles) representing various node types, connected by lines and arrows to demonstrate network relationships and traceability features. All prototypes are sketched on white paper with pen and pencil markings, showing participants' envisioned systems for qualitative data analysis and causal relationship visualization.}
    \label{fig:sketch}
\end{figure*}

\subsection{Findings}

\subsubsection{Current Practices for Exploring Causal Relationships and Generating Hypotheses}

Participants described their typical workflow as following a progression through multiple phases. 
The process begins with reading transcripts, as I3 noted: ``\textit{Often if you do interviews, you might have tens or even thousands of pages of text and maybe we have to read the whole script.}'' 
This initial reading phase is followed by \textit{keyword extraction}, as P2 explained: ``\textit{The second step, which is keyword extraction. We believe that basically looking for certain keywords will get rid of the [...] isomorphism of the words,}'' followed by ``\textit{coding keywords, which is essentially because even though we have keywords, certain keywords may express the same concepts.}'' 
The final step involves manually identifying connections between coded elements, with I1 describing: ``\textit{Then I start with identifying the relationships between those labels. So what are the causes? What are the consequences? What are the phenomena?}''
F2 similarly explained their goal to ``\textit{put things into reason versus decision, get the primary results from which influences which.}''

\subsubsection{Challenges in Causally Connecting Qualitative Data}

Participants identified several major challenges associated with generating hypotheses from qualitative data.
First, they found \textit{keyword coding} to be one of the most challenging aspects. 
P3 articulated this difficulty: ``\textit{We think the most difficult thing is the keyword coding because people use different ways of expressing similar concepts and actions.}''
P6 reinforced this concern about ``\textit{extracting the keywords}'' when ``\textit{multiple keywords can mean the same thing.}''
Second, participants acknowledged the challenge of distinguishing \textit{correlation} from \textit{causation} when identifying causal relationships, with this confusion potentially compromising their analysis (P7, P9). 
And third, the interpretive complexity of qualitative data posed additional difficulties.
P7 noted: ``\textit{[...] We need to understand the real meaning of the script [...] maybe if there is sarcasm or something like that, you can read behind the stories.}''
P10 similarly highlighted how cultural context shapes interpretation: ``\textit{There are some factors related to culture. So if we're from different cultural backgrounds, when we try to identify the keywords, we may be affected by our background and we may give different interpretations to the same word.}''

\subsubsection{Desired Features for Computational Causal Discovery Tools}

In response to these challenges, participants articulated specific visions for computational tools.
First, P5 described their vision for ``\textit{automatic word extraction for categories}'' because ``\textit{in the very initial stage, actually the most effort requiring stage [...] we need to categorize the different texts into different groups}'' and ``\textit{we hope that here it can save our time.}''
P11 envisioned tools that could ``\textit{extract the same meaning from multiple clusters [...] and automatically extract all the relevant keywords to something important.}''
F2 suggested a specific ``\textit{auto-complete function in the keyword part to prevent the AI from misunderstanding the data, as some of the phrases it provided were incomplete.}''

Second, participants also envisioned computational tools that could assist with causal relationship detection. 
P9 expressed this aspiration: ``\textit{We would like to have an ideal tool to help us determine whether the relationship is a causal or non-causal relationship.}''

Third, multi-level visualization emerged as essential for revealing broad patterns and enabling traceability to the source data.
F6 emphasized the need for visual solutions when dealing with large datasets: ``\textit{A popular feature that we need is visualization. So we're working with some very big data sets, and the graph could be very big, right? How can we visualize this [the data] in a meaningful way so that we can identify some patterns?}''
P2 similarly described their vision for integrating visual elements with source data: ``\textit{I think it would be very nice if you have a causal graph with the labels and nodes and the relationships between them. And you also have the instances of the original text associated with it [the node]. Then you might be able to do some more dynamic inference automatically from your causal graph.}''

The paper prototypes further revealed participants' preference for visual representations (Figure \ref{fig:sketch}).
One interesting observation was that directed graphs and networks featured different node shapes to represent various types of text or keywords. 
This visual approach was consistently paired with interactive functionality, as participants uniformly included clickable nodes that would enable traceability back to source data.

Finally, participants emphasized that computational tools should preserve human agency rather than fully automate the analytical process. 
I3 expressed this preference for collaborative systems: ``\textit{Humans should be in the loop for this tool.}''

\subsection{Design Goals}

Based on insights from the formative workshop, we formalize the following design goals (DG\#) and requirements (R\#) for our system:

\begin{itemize}
\item \textbf{[DG1] Characterize, categorize, and causally connect qualitative data in an interactive, user-controlled manner.} 
    \begin{itemize}
    \item \textbf{[R1]} The system should reduce the effort required to extract key information, conceptualize qualitative data, and identify the causal relationships.
    \item \textbf{[R2]} The system should allow users to modify and interpret its output.
    \end{itemize}
\item \textbf{[DG2] Illustrate, organize, and visualize qualitative data at multiple levels.} 
    \begin{itemize}
    \item \textbf{[R3]} The system should help users explore broad relational patterns and gain a general understanding.
    \item \textbf{[R4]} The system should provide access to details and source data as context.
    \end{itemize}
\end{itemize}

%% file: sections/4_system_design.tex
\section{System Design}
\label{sec:system}

Based on our design goals, we developed \texttt{QualCausal}, a system that helps users explore causal relationships and generate hypotheses using \colorbox{processbrown!20}{causal networks}.
Specifically, \texttt{QualCausal} automatically extracts \colorbox{nodeblue!20}{indicators} (DG1-R1) that users can curate, abstract, and organize into \colorbox{conceptorange!20}{concepts} (DG1-R2). 
The system then uses LLMs to detect \colorbox{edgegreen!20}{causal relationships} between indicators \cite{causal_nlp_egami_2022, causal_inference_feder_2022, causal_nlp_feuerriegel_2025} and form a causal network. 
The causal network construction process is shown in Figure \ref{fig:open_coding}.
Next, \texttt{QualCausal} uses an interactive directed graph interface with cascading views (i.e., \textit{Indicator View}, \textit{Concept View}, \textit{Node View}, and \textit{Details and Concepts Panel}) to visualize the causal network \cite{causal_viz_guo_2023}, enabling users to explore overall patterns (DG2-R3) and specific \colorbox{edgegreen!20}{causal relationships} in detail alongside the source text (DG2-R4). 
The causal network exploration interface is showcased in Figure \ref{fig:exploration}.
As introduced below, the system has two main features: a webpage for constructing \colorbox{processbrown!20}{causal networks} and a set of visual panels for exploring them.

\paragraph{\textcolor{darkred}{\textbf{Implementation.}}}
\label{sec:implementation}

\texttt{QualCausal} was initially prototyped in Figma and then developed using Vue.js, D3.js, and Django. 
The system leverages GPT-4.1 to power all LLM functionalities, including indicator extraction, concept creation, and causal relationship extraction. 
To facilitate easy access for participants, the system was deployed as a web application hosted on AWS infrastructure, with all participant input and output data securely stored on our AWS server without personal identifiers to ensure privacy protection.
Our system is open-sourced and available online.\footnote{\url{https://github.com/guguqiui/QualCausal}}

\begin{figure*}
    \centering
    \includegraphics[width=1\linewidth]{figures/figure_construction.jpg}
    \caption{Interface for causal network construction. (a) \textbf{Indicator Extraction}: The system automatically extracts indicators from the uploaded data (a.1) based on the user-provided research overview. Users can select text to edit indicators (a.2), type to modify them (a.3), or delete indicators (a.4). (b) \textbf{Concept Creation}: Users can create concepts to abstract indicators into meaningful theoretical constructs (b.1) and assign colors to them (b.2). Additionally, the system allows users to map indicators to the appropriate concepts (b.3) and save these mappings (b.4).}
    \Description{Screenshot of the Constructing Causal Network interface, where raw qualitative data are segmented into indicators and mapped to concepts. The left panel displays a table with three columns: original sentences, automatically detected indicators, and drop-down menus for assigning each indicator to a specific concept. Users can edit indicators by first clicking the pencil icon, and then either selecting text directly from the sentence to replace indicators or typing directly to modify them. Indicators can also be deleted using the trash icon. The right panel lists all available concepts, each with customizable colors, names, definitions, and references. Users can add new concepts using the green ``Add Concept'' button at the bottom. This stage enables researchers to iteratively refine and conceptualize their qualitative data before moving on to causal relationship extraction and visualization.}
    \label{fig:open_coding}
\end{figure*}

\subsection{Interactively Constructing Causal Networks (DG1)}
\label{sec:dg1}


Our system operationalizes Grounded Theory \cite{grounded_theory_larossa_2005} and engineers previous work on causal network construction from qualitative data \cite{causal_graph_tong_2024, causal_text_meng_2025} into a user-controlled workflow (Figure \ref{fig:open_coding}).

\subsubsection{Research Overview}

Based on insights from our formative study (Section \ref{sec:formative}), in order for the system to better process uploaded data, users must first provide a research overview (Figure \ref{fig:teaser}). 
This paragraph serves as a concise description of the study's focus and the type of qualitative data being analyzed. 
It is concatenated into the prompts for the LLM to perform indicator extraction \cite{causal_graph_tong_2024}, providing the necessary domain knowledge and context. 
Full prompts can be found in \textit{Supplementary Materials}.

\subsubsection{Indicator Extraction/Node Formation}
\label{sec:indicator}

\texttt{QualCausal} automatically identifies \colorbox{nodeblue!20}{indicators} and allows users to validate or/and refine them. 
These \colorbox{nodeblue!20}{indicators} are visualized as nodes in the causal network.

The system parses the uploaded dataset line by line and extracts \colorbox{nodeblue!20}{indicators} based on the research overview. 
The extracted \colorbox{nodeblue!20}{indicators} are displayed in a tabular format with source sentences on the left and their corresponding indicators on the right, allowing users to trace each indicator back to its source context.
Users can review all \colorbox{nodeblue!20}{indicators} through a paginated table interface (Figure \ref{fig:open_coding} (a.1)), record analytical memos, and perform two curation operations. 
First, users can modify any \colorbox{nodeblue!20}{indicators} that they consider imprecise, either by highlighting text segments from the original sentence (Figure \ref{fig:open_coding} (a.2)), which will automatically populate the edit field, or by rewriting the indicator manually (Figure \ref{fig:open_coding} (a.3)).
Additionally, users can delete \colorbox{nodeblue!20}{indicators} (Figure \ref{fig:open_coding} (a.4)) they find redundant or irrelevant to their research focus.

\subsubsection{Concept Creation/Node Type Assignment}
\label{sec:concept}

\texttt{QualCausal} allows users to categorize and conceptualize qualitative data by creating \colorbox{conceptorange!20}{concepts} (Figure \ref{fig:open_coding} (b.1)) and supports automatic, deductive mapping of \colorbox{nodeblue!20}{indicators} to \colorbox{conceptorange!20}{concepts} for determining node types in the causal network.

Users can allocate colors to each concept via the concept-creation interface located on the right side of the system (Figure \ref{fig:open_coding} (b.2)).
Each concept consists of three fields: \textit{name} (concept label), \textit{definition} (conceptual description), and \textit{reference(s)} (i.e., supporting example(s) from the dataset) \cite{hci_qual_lazar_2017, grounded_theory_larossa_2005}. 
The \textit{name} and \textit{definition} fields require manual input from users, while the \textit{reference(s)} field can be populated in two ways: users can manually type relevant examples they consider appropriate, or the system automatically populates this field when users assign \colorbox{conceptorange!20}{concepts} to specific \colorbox{nodeblue!20}{indicators} using the dropdown menu adjacent to each indicator in the table (Figure \ref{fig:open_coding} (b.3)). 
The \textit{reference(s)} field stores up to three examples to prevent bias toward reference-rich concepts during automated mapping (Figure \ref{fig:open_coding} (b.4)). 

Once the concept list is finalized, the system reads it and incorporates its three fields (i.e., \textit{name}, \textit{definition}, and \textit{reference(s)}) into a structured prompt (see \textit{Supplementary Materials}).
This prompt serves as instructions for automatically assigning \colorbox{conceptorange!20}{concepts} to unmapped \colorbox{nodeblue!20}{indicators}, thereby completing the node type assignment for the entire causal network.

\subsubsection{Causal Relationship Extraction/Edge Generation}
\label{sec:causal}

In the final step of building the causal network, \texttt{QualCausal} performs automated causal discovery \cite{causal_extract_liu_2025}. 
First, it generates all possible combinations of indicator pairs within each sentence. 
These pairs, along with their associated \colorbox{conceptorange!20}{concepts} and the original sentence for context, are submitted to LLMs.
The system then prompts the LLM to determine whether \colorbox{edgegreen!20}{causal relationships} exist between the indicator pairs and, if so, which direction the causation flows, thereby forming directed edges for the causal network. 
Specifically, the LLM performs a three-way classification task, categorizing each pair as ``\textit{Indicator 1 causes Indicator 2},'' ``\textit{Indicator 2 causes Indicator 1},'' or ``\textit{no direct causal relationship}'' based on both explicit causal markers (e.g., ``\textit{because},'' ``therefore'') and implicit causal relationships inferred from context \cite{causal_extract_antonucci_2023}. 
We adapted the zero-shot prompting strategy from previous work on pairwise causal discovery \cite{causal_extract_antonucci_2023, causal_extraction_kiciman_2023, causal_extract_wan_2024}, with the full prompt available in \textit{Supplementary Materials}.

After identifying \colorbox{edgegreen!20}{causal relationships}, the system merges similar and potentially redundant indicators based on their semantic meaning and context (see \textit{Supplementary Materials} for details) to produce the final \colorbox{processbrown!20}{causal networks}.

\begin{figure*} 
\centering 
\includegraphics[width=1\linewidth]{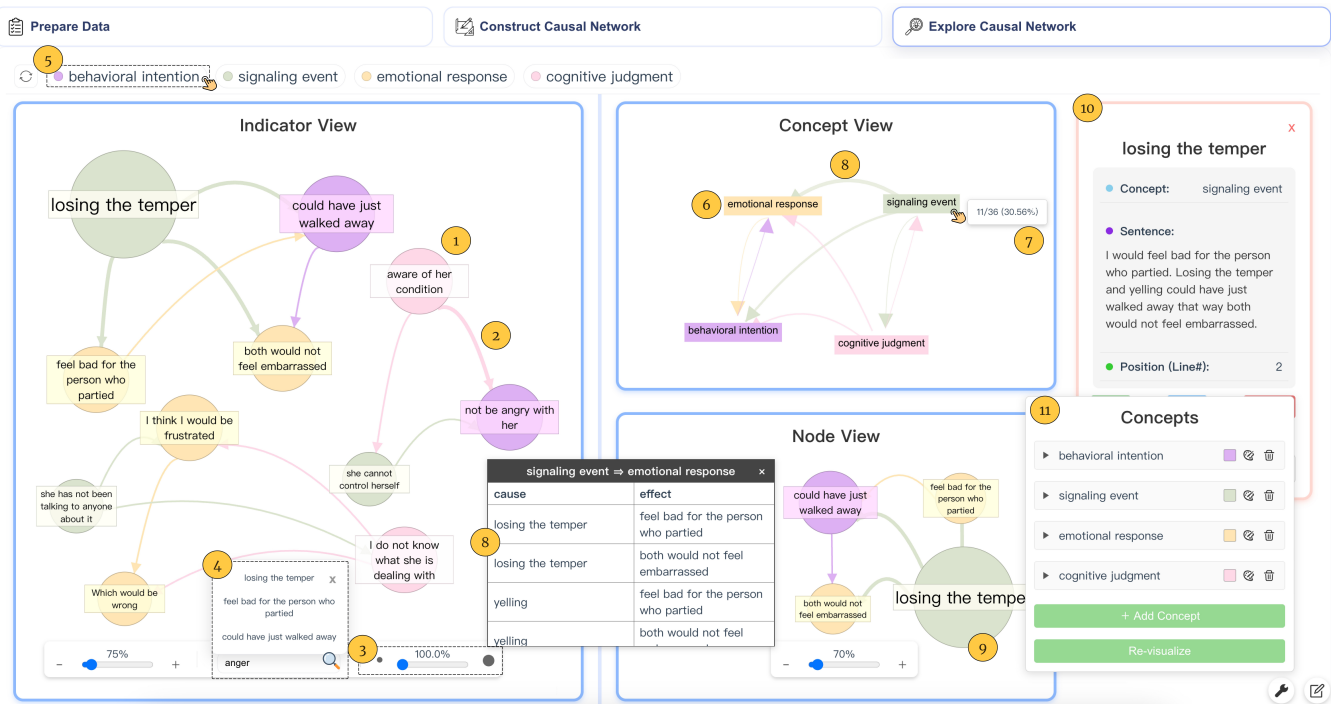} 
\caption{Interface for causal network exploration with multiple coordinated views. \textbf{Indicator View}: causal network display where nodes represent indicators and edges denote causal relationships. Interactions include connectivity-based node sizing, slider control, semantic search, and legend-based filtering (\textcircled{1}-\textcircled{5}). \textbf{Concept View}: Abstracted conceptual model showing user-created concepts as colored blocks connected by weighted edges. Interactions include hovering for percentages and clicking edges for supporting indicator pairs (\textcircled{6}-\textcircled{8}). \textbf{Node View}: Localized subnetwork centered on selected indicators (\textcircled{9}). \textbf{Details and Concepts Panel}: Sidebar for viewing context and editing concepts with re-visualization (\textcircled{10}-\textcircled{11}).}
\Description{Screenshot of the Exploring Causal Network interface, which provides four coordinated views for analyzing qualitative data. The Indicator View on the left visualizes individual indicators as nodes and their causal relationships as edges. Node size reflects connectivity, and a slider allows filtering to focus on the most connected nodes. Users can search for indicators by keywords and highlight nodes by selecting concept colors from the legend. The Concept View in the top center shows higher-level relationships between concepts, with edge thickness indicating the strength of connections. The Node View at the bottom center displays the localized subnetwork of a selected indicator, helping researchers focus on its immediate context. The Details and Concepts Panel on the right shows the original sentence containing the selected indicator and allows researchers to edit concept details, manage colors, and re-visualize the network. This integrated interface supports exploration from detailed indicator-level analysis to abstract conceptual understanding.}
\label{fig:exploration} 
\end{figure*}

\subsection{Exploring Causal Networks through Multi-view Visualization (DG2)}
\label{sec:dg2}

After constructing a causal network of nodes (\colorbox{nodeblue!20}{indicators}) and their conceptual labels (\colorbox{conceptorange!20}{concepts}) as well as edges (\colorbox{edgegreen!20}{causal relationships}), \texttt{QualCausal} visualizes the network through multiple coordinated views. 
These views allow users to explore and make sense of their qualitative data from different perspectives and levels of detail, as shown in Figure \ref{fig:exploration}. 
The usage scenario presented below:

\begin{quote}
\textit{Cynthia is a researcher studying how mental-health stigma forms using interview data. 
After creating her causal network with \texttt{QualCausal}, she begins by exploring the Indicator View, zooming and dragging to get an overview. 
She adjusts the slider to display about 20 nodes for clarity and notices that purple nodes dominate the display. 
Clicking the purple box in the legend highlights all indicators with the conceptual label ``behavioral intention,'' revealing that they comprise 25\% of her data.}

\textit{Moving to the Concept View, Cynthia sees that the thickest edge pointing to ``behavioral intention'' comes from ``cognitive judgment.'' 
Hovering over the edge reveals that it represents 15\% of the concept-level causal relationships. 
Clicking the edge opens a detailed table showing specific indicator pairs, including ``they are unpredictable'' leading to ``keep away from them.''}

\textit{Intrigued by unpredictability, Cynthia returns to the Indicator View and searches for it. 
\texttt{QualCausal} returns ``unpredictability is dangerous,'' so she clicks to display the subnetwork in Node View. 
She finds that this cognitive judgment also leads to ``may feel angry.''
To better understand this edge, she checks the original sentence in the Details and Concepts Panel and reads: ``I probably may feel angry because I think their unpredictability is dangerous since you never know when they lose control.''}

\textit{During iterative exploration, Cynthia adds the new concept ``personality traits'' and re-visualizes the causal network. 
She concludes that the causal relationship in which unpredictability triggers both anger and social distancing is reasonable and interesting -- a hypothesis she plans to test in future research.}
\end{quote}

\subsubsection{Indicator View}

The \textit{Indicator View} helps users identify broad patterns and gain an overall sense of their data (Figure \ref{fig:exploration} \textcircled{1}-\textcircled{2}). 
Node colors represent the \colorbox{conceptorange!20}{concepts} assigned to each indicator, and node sizes vary based on connectivity -- \colorbox{nodeblue!20}{indicators} with more incoming and outgoing causal relationships appear as larger nodes, highlighting their potential importance in the causal network.
This view is interactive: users can zoom in and out, reposition nodes by dragging them, delete unwanted nodes, and edit both the indicator and its assigned concept. 
The default layout uses a force-directed algorithm to organize the network spatially.

We implemented several filtering mechanisms to facilitate exploration when the data volume is high and the causal network is large.
First, a slider control (Figure \ref{fig:exploration} \textcircled{3}) allows users to display only a certain percentage of nodes, keeping the most connected indicators (i.e., larger nodes) visible. 
Second, a search box (Figure \ref{fig:exploration} \textcircled{4}) helps users locate specific nodes based on semantic similarity. 
And third, the legend displays colored boxes (Figure \ref{fig:exploration} \textcircled{5}) representing each concept, and clicking on any of these colored boxes retains only the nodes of that single color while dimming all others. 
This feature helps users focus their attention on \colorbox{nodeblue!20}{indicators} within a single concept.

\paragraph{\textcolor{darkred}{\textbf{Interaction with Other Views.}}}

Clicking on a node or selecting an indicator from the semantic search dropdown triggers coordinated updates across all views: the \textit{Concept View} highlights the colored block of the corresponding concept while dimming others, the \textit{Node View} displays the connected component of the selected indicator, and the \textit{Details and Concepts Panel} shows the corresponding original sentence. 
When users click on an edge in the \textit{Indicator View}, the \textit{Concept View} highlights the corresponding concept-level edge between the two concepts to which the connected indicators belong, while the \textit{Node View} presents the subnetwork containing that specific causal relationship.

\subsubsection{Concept View}

The \textit{Concept View} helps users discover causal relationships between \colorbox{conceptorange!20}{concepts} and generate hypotheses by abstracting the indicator-level causal network into a higher-level conceptual model. 
It displays colored blocks representing \colorbox{conceptorange!20}{concepts} (Figure \ref{fig:exploration} \textcircled{6}) connected by directed edges depicting \colorbox{edgegreen!20}{causal relationships}.

This view is computationally derived from the \textit{Indicator View}, where multiple causal relationships between \colorbox{nodeblue!20}{indicators} assigned to the same pair of \colorbox{conceptorange!20}{concepts} and flowing in the same direction are consolidated into a single concept-level edge with accumulated \textit{weight}.
Edge thickness encodes the \textit{weight} of \colorbox{edgegreen!20}{causal relationships}, with thicker edges indicating that more indicator pairs between those two \colorbox{conceptorange!20}{concepts} have causal relationships in the data.

The interface provides rich interactive features for exploration. 
It allows users to organize concepts according to their analytical needs, such as arranging them in layers.
Besides, hovering over a colored block displays the percentage of \colorbox{nodeblue!20}{indicators} assigned to that specific concept (Figure \ref{fig:exploration} \textcircled{7}), and hovering over an edge shows the percentage of that concept-level causal relationships.
Clicking on any concept-level edge opens a detailed table (Figure \ref{fig:exploration} \textcircled{8}) displaying all the specific indicator pairs that contribute to that causal relationship.

\paragraph{\textcolor{darkred}{\textbf{Interaction with Other Views.}}}

Clicking on a colored block highlights all the \colorbox{nodeblue!20}{indicators} belonging to that concept in both the \textit{Indicator View} and \textit{Node view}. 
Similarly, clicking on a concept-level edge highlights all the underlying indicator-to-indicator causal relationships that contribute to that edge in both views. 

\subsubsection{Node View}

The \textit{Node View} displays a localized causal subnetwork centered on a focal indicator, which helps users analyze nuances when investigating specific \colorbox{nodeblue!20}{indicators} or finding evidence of a particular causal relationship. 
This view operates with similar coordinated interactions as the \textit{Indicator View}, updating dynamically when users select nodes or edges from other views to show the relevant local network structure (Figure \ref{fig:exploration} \textcircled{9}).

\subsubsection{Details and Concepts Panel}

Beyond the three visual panels, the \textit{Details and Concepts Panel} on the right side of the interface serves dual purposes: 1) providing access to original text sources and 2) offering controls for managing concepts and re-visualizing \colorbox{processbrown!20}{causal networks}. 
For each selected indicator, the panel displays its assigned concept, the original sentence from which it was extracted, and its position within the source data (Figure \ref{fig:exploration} \textcircled{10}). 
The panel also allows users to edit existing concepts, create new ones, and refresh and re-visualize the causal network across all views after making changes (Figure \ref{fig:exploration} \textcircled{11}).

\subsection{Technical Evaluation}
\label{sec:technical}

\begin{table*}[t]
\centering
\caption{Technical evaluation of \textit{Indicator Extraction}, \textit{Concept Creation}, and \textit{Causal Relationship Extraction} across SemEval 2010 Task 8 and MHStigmaInterview-20 datasets.}
\Description{Table presenting the technical evaluation of Indicator Extraction, Concept Creation, and Causal Relationship Extraction across SemEval 2010 Task 8 and MHStigmaInterview-20 datasets. The table compares three methods: Co-occurrence networks, Causal-cue heuristics, and Our system. For SemEval 2010 Task 8 (Causal Relationship Extraction), Co-occurrence networks achieve 38.00\% Precision/Directionality Accuracy and 100.00\% Recall; Causal-cue heuristics achieve 88.89\% Precision and 8.00\% Recall; Our system achieves 100.00\% Precision and 85.00\% Recall. For MHStigmaInterview-20, the ground truth data has Inter-Annotator Agreements of 0.93 (Positive Specific Agreement) for Indicator Extraction, 0.92 (Cohen's Kappa) for Concept Creation, and 0.92 (Positive Specific Agreement) for Causal Relationship Extraction. In the Indicator Extraction task, Our system achieves 86.46\% Precision and 94.32\% Recall. In the Concept Creation task, Our system achieves 66.39\% Precision, 76.96\% Recall, and 70.83\% Accuracy. In the Causal Relationship Extraction task, Co-occurrence networks achieve 38.52\% Precision, 72.31\% Recall, and 72.31\% Directionality Accuracy; Causal-cue heuristics achieve 22.58\% Precision, 10.77\% Recall, and 87.50\% Directionality Accuracy; Our system achieves 80.60\% Precision, 83.08\% Recall, and 98.18\% Directionality Accuracy.}
\label{tab:tech_eval}
\resizebox{\textwidth}{!}{%
\begin{tabular}{l cc ccc cccc cccc}
\toprule
& \multicolumn{2}{c}{\textbf{SemEval 2010 Task 8}} & \multicolumn{11}{c}{\textbf{MHStigmaInterview-20}} \\
\cmidrule(lr){2-3} \cmidrule(lr){4-14}
& \multicolumn{2}{c}{Causal Relationship Extraction} & \multicolumn{3}{c}{Indicator Extraction} & \multicolumn{4}{c}{Concept Creation} & \multicolumn{4}{c}{Causal Relationship Extraction} \\
\cmidrule(lr){2-3} \cmidrule(lr){4-6} \cmidrule(lr){7-10} \cmidrule(lr){11-14}
\textbf{Method} & Prec / DirAcc & Rec & Prec & Rec & \begin{tabular}{@{}c@{}}IAA \\ (PSA)\end{tabular} & Prec & Rec & Acc & \begin{tabular}{@{}c@{}}IAA \\ ($\kappa$)\end{tabular} & Prec & Rec & DirAcc & \begin{tabular}{@{}c@{}}IAA \\ (PSA)\end{tabular} \\
\midrule
Co-occurrence networks & $38.00\%$ & $100.00\%$ & - & - & \multirow{3}{*}{0.93} & - & - & - & \multirow{3}{*}{0.92} & $38.52\%$ & $72.31\%$ & $72.31\%$ & \multirow{3}{*}{0.92} \\
Causal-cue heuristics  & $88.89\%$ & $8.00\%$ & - & - &  & - & - & - & & $22.58\%$ & $10.77\%$ & $87.50\%$ & \\
Our system             & $100.00\%$ & $85.00\%$ & $86.46\%$ & $94.32\%$ & & $66.39\%$ & $76.96\%$ & $70.83\%$ & & $80.60\%$ & $83.08\%$ & $98.18\%$ & \\
\bottomrule
\end{tabular}%
}
\end{table*}

We quantitatively evaluated the pipeline our system uses to construct causal networks (Section \ref{sec:dg1}). 
Specifically, we report the performance and execution time for indicator extraction (Section \ref{sec:tech_indicator}), concept creation (Section \ref{sec:tech_concept}), and causal relationship extraction (Section \ref{sec:tech_causal}) below and in Table \ref{tab:tech_eval}.

\subsubsection{Evaluating Indicator Extraction}
\label{sec:tech_indicator}

To assess how well our system extracts indicators (Section \ref{sec:indicator}), we first built a dataset (MHStigmaInterview-20) containing 20 responses ($M=37.00$, $SD=16.09$ words per sentence) sampled from a public interview corpus on mental-health stigma (MHStigmaInterview) \cite{sample_data_meng_2025}, which we determined was suitable for using computational tools for qualitative data analysis and causal discovery after consulting with experts on our team. 
These responses addressed the question ``\textit{If you were in the process of selecting a tenant for your home, would you feel comfortable entrusting it to someone like Avery?}'' 
A sample response read, ``\textit{Not at all. Unforeseen events and not knowing how the person will react on other days. Avery might have mood swings and cannot control them.}''

Two researchers independently annotated the indicators, with an inter-annotator agreement (positive specific agreement, PSA \cite{psa_hripcsak_2005}) of $0.93$.\footnote{When comparing two sets of annotator-extracted indicators or validating system-extracted indicators, we considered them a match even if the wording was not exactly the same, as long as the core meaning remained the same. For example, ``\textit{lead to my property depreciating in value}'' and ``\textit{may lead to my property depreciating in value}'' were considered as a match because both capture the same underlying concern about property value loss.}
We then used an expert-written research overview, ``\textit{the study explores the causal factors that cause stigma against people with mental illness},'' to guide the indicator extraction. 
The system extracted 96 indicators from the 20 responses with an average execution time of $M=39.21$ seconds ($SD=1.80$) over five runs,\footnote{Reported execution times were measured on a MacBook Pro (Apple M1 Pro chip, 16GB RAM) running macOS 14.5, using Python 3.10.9 and OpenAI SDK 1.58.1. This configuration applies to all reported times in the paper.} achieving a precision of $86.46\%$ and a recall of $94.32\%$. 
This shows that our system extracts indicators that are generally accurate and comprehensive in their coverage of the source data.

\subsubsection{Evaluating Concept Creation}
\label{sec:tech_concept}

We used the same MHStigmaInterview-20 dataset to assess how well our system automatically maps indicators to concepts (Section \ref{sec:concept}).
We drew the \textit{name} and \textit{definition} of four concepts from the attribution theory of mental-health stigma \cite{attribution_theory_corrigan_2000} (i.e., signaling event, cognitive mediator, affective response, and behavioral reaction), with their \textit{reference(s)} field left empty.

Our system mapped 24 indicators to signaling events (25.00\%), 48 to cognitive mediators (50.00\%), 8 to affective responses (8.33\%), and 16 to behavioral reactions (16.67\%), with an average processing time of $M=123.96$ seconds ($SD=10.84$) across five independent runs. 
Two researchers independently validated these mappings, achieving an inter-annotator agreement (Cohen's $\kappa$ \cite{kappa_mchugh_2012}) of 0.92. 
Comparing the consolidated annotations with the system's output yielded a precision of $66.39\%$, a recall of $76.96\%$, and an accuracy of $70.83\%$. 
This demonstrates that our system could partly support abstracting indicators into higher-level conceptual categories.

\subsubsection{Evaluating Causal Relationship Extraction}
\label{sec:tech_causal}


\paragraph{\textcolor{darkred}{\textbf{Datasets.}}}

We evaluated causal relationship extraction (Section \ref{sec:causal}) on two datasets. 
The first was the same MHStigmaInterview-20 dataset described above, where two researchers independently annotated causal relationships ($M=3.25$ per sentence) with an inter-annotator agreement (PSA \cite{psa_hripcsak_2005}) of 0.92. 

The second was a random sample of 100 sentences ($M=19.34$, $SD=7.55$ words per sentence) from the widely used SemEval 2010 Task 8 dataset,\footnote{\url{https://huggingface.co/datasets/SemEvalWorkshop/sem_eval_2010_task_8}} specifically from the 659 sentences annotated as having cause-and-effect relationships. 
This dataset has annotated pairs of words in each sentence that hold causal relationships (inter-annotator agreement of 79\% \cite{semeval_dataset_hendrickx_2010}). 
An example is ``\textit{The \texttt{<e1>}gaps\texttt{</e1>} in the rings are caused by \texttt{<e2>}resonance\texttt{</e2>} between the particles in the rings and the moons orbiting nearby.}'' 
We should note that the \textit{annotated words} in this dataset differ from \textit{indicators} in qualitative data analysis, as they are not abstracted into \textit{concepts} or intended for hypothesis generation. 
We use this dataset to confirm whether our system can extract ground-truth causal relationships and to establish external validity.

\paragraph{\textcolor{darkred}{\textbf{Baselines.}}}

We compared our system against two commonly used methods in existing systems for exploring causal relationships: 1) \textbf{co-occurrence networks} \cite{visual_cooccurrence_schwab_2024, visual_cooccurrence_pokorny_2018, visual_cooccurrence_andrienko_2020, visual_cooccurrence_chandrasegaran_2017}, which determine causal direction between any two adjacent indicators by treating earlier ones as causes and later ones as effects, and 2) \textbf{causal-cue heuristics} (used in NVivo's query features) \cite{causal_nlp_altenberg_1984, causal_heu_khoo_1998, causal_nlp_newberry_2024}, which detect causal relationships through causal keyword lists (e.g., \textit{so}, \textit{therefore}) and linguistic patterns (e.g., ``\textit{\texttt{[effect]} is the result of \texttt{[cause]}}'') \cite{causal_heu_khoo_1998}, implemented using regular expressions. 

\paragraph{\textcolor{darkred}{\textbf{Results.}}}

Our system outperformed both baselines across the two datasets. 
On MHStigmaInterview-20, our system extracted 67 causal relationships ($M=3.35$ per sentence) with an average processing time of $M=226.85$ seconds ($SD=12.80$) over five trials, achieving a precision of $80.60\%$, a recall of $83.08\%$, and a directionality accuracy of $98.18\%$, compared to co-occurrence networks ($M=6.10$ per sentence, $Prec=38.52\%$, $Rec=72.31\%$, $DirAcc=72.31\%$) and causal-cue heuristics ($M=1.55$ per sentence, $Prec=22.58\%$, $Rec=10.77\%$, $DirAcc=87.50\%$). 
For the SemEval 2010 Task 8 subsample, our system extracted 85 causal relationships ($M=0.85$ per sentence) in $M=186.18$ seconds ($SD=0.69$), resulting in a precision/directionality accuracy of $100.00\%$ and a recall of $85.00\%$, while co-occurrence networks ($M=1.00$ per sentence, $Prec/DirAcc=38.00\%$, $Rec=100.00\%$) and causal-cue heuristics ($M=0.10$ per sentence, $Prec/DirAcc=88.89\%$, $Rec=8.00\%$) performed less well. 
Upon inspection, we found that co-occurrence networks tend to reverse causal directions (hence the relatively low precision), and causal-cue heuristics struggle to capture implicit causal relationships (hence the relatively low recall), problems that our system partially mitigates through semantic and contextual understanding.

%% file: sections/5_feedback_study.tex
\section{Feedback Study Design}
\label{sec:feedback}


To understand whether \texttt{QualCausal} helps users discover meaningful causal relationships, how users might apply computational tools in their research, and what concerns they have about such tools, we conducted a qualitative feedback study by interviewing users.

\subsection{Participants}

We recruited participants through formative study, social media channels, and personal networks.
Of the 15 participants (6 men, 9 women in their twenties and early forties), five were faculty members, four were industrial researchers, and six were Ph.D. students (Table \ref{tab:participants}), eight of whom had participated in the formative study. 
We balanced the recruitment of participants with prior experience using computational tools for qualitative data analysis ($n=7$) with those who were interested but had limited experience ($n=8$, based on recruitment questionnaire responses). 
We compensated participants US\$16.5 on average via local currency. 
The study received approval from our institutional ethics review committee.

\subsection{Procedure}

The one-hour in-person interview (Figure \ref{fig:study_design} (b)) began with obtaining the participants' permission to make an audio recording. 
Then, we introduced them to the MHStigmaInterview-20 dataset (refer to Section \ref{sec:technical}) and walked them through our system \cite{walkthrough_light_2018}, explaining its main features. 
We used the same dataset for all participants to control for potential bias arising from differences in data familiarity.
An example research scenario was set for them to explore how mental-health stigma forms, specifically investigating the causes of stigma and the underlying mechanisms.
With step-by-step guidance, we asked the participants to think aloud \cite{think_aloud_hertzum_2024} as they completed two tasks: 1) collaborating with \texttt{QualCausal} to construct the causal network interactively, and 2) navigating the generated network using multiple visual panels.

After the hands-on session, we interviewed participants about their experiences with our system, focusing on 1) what causal relationships or potential hypotheses they identified using \texttt{QualCausal}, 2) how they used the system to derive these findings, 3) which system features they found most useful and why, and 4) what additional features they wished for and why.
We then broadened the interview to a more general discussion, asking 1) what they saw as the main benefits or potential applications of such a system and how they might use it, and 2) what concerns or risks they associated with using computational tools for causal discovery in qualitative data and why.

We audio-recorded and transcribed the sessions using Microsoft Word, then analyzed the transcripts using the same thematic analysis approach \cite{thematic_analysis_braun_2006} as the formative study.



\section{Feedback Study Findings}

\subsection{How Did Participants Experience Interactive Causal Network Construction?}

\subsubsection{In what ways did automatic indicator extraction contribute to qualitative data analysis?}

Participants identified several benefits of the extracted indicators, such as increasing efficiency (DG1-R1) and directing their attention to relevant content within lengthy passages. 
F2 stated: ``\textit{It [indicator extraction] is really fast, [and it] lets me know where to focus. For very long sentences, it [indicators] can tell me which parts I actually need to pay attention to.}''
Chunking sentences into indicators also reduced cognitive load; P12 observed that ``\textit{although the total word count hasn't changed, the reading approach has changed. [...] It's all 20 words, but reading 20 words all at once versus reading them in chunks of 5 feels different.}''
In addition, the indicators served as references and analytical baselines, which F2 described as ``\textit{a baseline that I can use as a starting point [...] It provides a reference and helps me get a big picture of the data. It's like an initial idea that I can then refine.}''
Lastly, the comprehensive nature of the indicators was appreciated by participants, with P4 noting that ``\textit{[...] It's good redundancy. More [indicators] for me to subtract is definitely better than fewer to add,}'' and I5 emphasizing that, ``\textit{unlike with NVivo, [...] which sometimes I feel FOMO, here I know it extracted everything, nothing important would be missed.}''

Despite these benefits, participants were concerned that automatic indicator extraction required high data cleanliness, and that off-topic content in first-hand data could not be removed, as P8 reflected that ``\textit{some people during interviews [...] might answer one of your questions, then ramble about their own stuff for a bit, then answer another question, and the machine processes everything.}''
Besides, participants recognized their tendency toward over-reliance. 
P9 candidly admitted: ``\textit{After it extracts indicators, I naturally don't look at it [the original text] anymore [...] I feel like I get lazy. I read through it [indicators] once briefly, then I want to classify directly rather than revise to make it [indicators] more accurate.}''

\subsubsection{How did participants engage with concept creation and mapping?}

Participants appreciated that the system maintained their autonomy over concept creation and mapping (DG1-R2) for several reasons.
First, the partial automation kept them engaged with the data by serving as cognitive forcing, as I5 said: ``\textit{I think maybe it [system] is forcing me to use my brain. Maybe for at least the first 100 codes, you can't use that AI one-click fill function -- you have to force me to think about it, but later, maybe I want to just click and get it [full results].}''
Second, participants recognized the risk of error propagation, as F1 explained: ``\textit{The common problem is the whole garbage in, garbage out [...] just making sure that you are not putting something garbage in there [...] I think researchers should always be hands-on to a certain extent with their data.}''
Third, participants noted that their usual practice of creating only informal concept descriptions necessitated human involvement.
P12 elaborated on this: ``\textit{If you suddenly tell me I need to give each concept a reference [...] I have to explain what the name means, what the definition is, what the reference is. But actually, I might just call everything in a single word in this stage.}''
Beyond practical assistance, participants valued the social aspect of working with the system.
I4 remarked: ``\textit{It's a bit of emotional support from the AI, as if there are two people coding together. As researchers, we're quite lonely, right? So there's somebody coding and you just agree or disagree -- it's like a partnership.}''

In contrast, participants conveyed concerns about how computational tools might alter their epistemological frameworks and methodological practices. 
P9 articulated a shift from exploring raw data to confirming the system's output: ``\textit{It turns interpretation into a classification task [...] now it [the indicator] becomes the signal that most attracts my attention, and to complete this classification task, I look for the thing [concepts] from here [indicators], and if this information can't satisfy me, then I go back [to the original text] [...] all the concepts I'm proposing now come from indicators, not from sentences.}''
P9 also noticed how the system changed analytical priorities: ``\textit{What I should do first when I see this data is read through everything, starting from my initial impression, because there will definitely be things that leave deep impressions [...] But now what I'm doing is going from the first indicator, second indicator, one by one.}''

\subsubsection{What were participants' perceptions of automatically generated causal relationships?}

Participants found that automated causal relationship extraction could address one of the most time-consuming parts of causal discovery from qualitative data (DG1-R1), as I2 explained: ``\textit{I don't think we can do this level of coding for all the transcripts [...] If you do it via Excel, that will be insane. So a lot of this judgment of the semantics [...] it's all based on the researcher's impression [...] I would have already done open coding and never done the axial coding. And then the relationship and drawing of the connections will be based on my understanding of all the transcripts [...], but not to this level of detail.}''
Furthermore, this automation helped participants overcome analytical inertia. 
I4 noted: ``\textit{I would appreciate it doing the connection for me, then I can agree or disagree [...] better than doing everything myself because there's an inertia to get started.}''
F5 similarly described the confidence that the extracted causal relationships could provide: ``\textit{It gives me an initial [framework] [...] sometimes you get overwhelmed with so much data. Then it [the system] gives you the initial [framework], and you realize you're both on the right page. When you do your work, it's faster and you're more confident.}''
Besides, F4 appreciated how system-generated causal relationships could reveal new ways of interpreting the data: ``\textit{I might use this to help me interpret data differently because the way I see different concept relationships might be different based on the system.}''

However, participants also raised several concerns. 
P9 highlighted the need to position causal relationships as exploratory rather than definitive: ``\textit{You're making a tool, don't think so complexly. Causality isn't your problem to solve, you just need to frame it as potential causal relationships [...] No qualitative research dares claim they found causation, and even if they did, no quantitative researcher would believe them.}''
Participants also emphasized that automated causal discovery requires deep negotiation with the system so that users can convey their own positionality. 
F4 commented: ``\textit{If you can feed the model your own values, your own knowledge, your own way of interpreting the data [...] but I'm not sure if machines have those kinds of values and whether we can really teach such deep complex perspectives to AI just through text.}''
Additional concerns emerged regarding the separation of categorization and connection phases, as P12 observed: ``\textit{It [system] supports categorization, but when people read, they're already starting to think about the causal relationships [...] For very large datasets, this kind of collaboration makes sense because it makes the whole project more manageable. But if it's very small, it's something I can do in one step -- I don't need to split it into two steps.}''

\subsection{How Did Participants Explore Causal Networks through Visualization?}

\subsubsection{How did participants use causal networks in the Indicator View?}

Interestingly, participants perceived the \textit{Indicator View} as helpful for both analyzing data and demonstrating and communicating findings. 
For instance, P10 appreciated the intuitive nature of causal network visualization: ``\textit{Compared to traditional coding or analysis methods, this is more intuitive [...] especially if the article is aimed at the public rather than people in the field, it would be much easier to understand.}''
Specifically, participants envisioned practical use cases. 
P4 noted the potential for publication use: ``\textit{If this tool eventually goes to market [...] and the copyright allows researchers to use them, that would be great, because I can imagine many people would want to put these figures in their articles [...] Many researchers are not good at making figures.}''
F5 likewise mentioned: ``\textit{Maybe if you're presenting to stakeholders or doing your thesis defense, it may be helpful.}''
Furthermore, the \textit{Indicator View} was particularly valuable when used in conjunction with other views to break down analytical complexity.
I3 observed that ``\textit{this Indicator [View is] quite useful, especially when you combine it [with] Concept View and Node View, which is very good. That can help address the complexity because if you put everything together, then you get lost.}''

Conversely, participants expressed concerns that the comprehensiveness of causal network visualization could potentially be overwhelming. 
F5 illustrated the challenge: ``\textit{If there's a lot of data, I really think it's going to be like you can't see the whole network on the screen [...] So you want to be able to delete some so that you can just show the main story. [...] I personally think main story is better, but some people may prefer the comprehensive story [...] For me, when I look at it, if I get confused, I won't look at it anymore. If the story is too complex, I can't [process it].}''

\subsubsection{How did the Concept View support causal discovery and hypothesis generation?}

We found that most participants could derive reasonable and meaningful causal relationships and hypotheses from the \textit{Concept View} (DG2-R3).
For instance, F1 interpreted the causal relationship shown in the \textit{Concept View}, explaining that ``\textit{it shows emotional illness affects troublemaker, so there might be a hypothesis that people would transform emotional instability or mental disorder into behavioral trouble.}'' 
Meanwhile, P1 affirmed the broader research value, stating that ``\textit{as long as these relations appear in qualitative findings, quantitative people can directly borrow these [concepts and causal relationships] from previous research.}''
Besides, the \textit{Concept View} interface, as F3 praised, ``\textit{is clearly showing faster, and is pretty easy to navigate and see what's all connected [...] we cannot manually visualize how these are connected together.}''
Specifically, the edge thickness and related numbers were useful for determining analytical priorities based on the strength of causal relationships, as I2 shared: ``\textit{AI does help -- 13 out of 40 have this relationship. It [edge weights] would help to offer some priority insights into what the prominent relationships are. [...] I'm very comfortable with numbers. You give me numbers justifying why certain relationships [exist], why certain concepts relate to each other. Love it.}''
I5 added: ``\textit{With manual [analysis], I have to look at them and think about what story they can tell. But this [edge weights] somehow gives me directionality and tells me which is most important.}''

Yet, some participants raised questions about discrepancies between the numbers indicating the strength of causal relationships and the underlying philosophy of qualitative data analysis, as P12 articulated: ``\textit{Qualitative researchers sometimes don't care if you're the majority. As long as you exist and I see you, I need to see you [...] The charm and meaning of qualitative [research] lies in the meaning-making afterward. We do this with humanistic care. We do social justice.}''
F4 elaborated on this methodological tension: ``\textit{Usually when I do grounded theory, I don't count each code numerically [...] I focus more on axial coding, which is more subjective [...] This [system] is based on numbers, so maybe [it's] more [like] a supplementary tool rather than using this tool to answer my research question.}''
Additionally, participants worried about potential information overload and misleadingness. 
P10 pointed out that ``\textit{it seems like it [system] connected everything that could be connected [...] I feel like it might be a bit too much,}'' while P4 observed: ``\textit{If two concepts have relationships where A points to B and B also points to A, both will be shown in this graph, but I think it would be more helpful if it just showed one [undirected edge] without arrows [...] Just tell me how related these two concepts are, but don't bias my judgment by telling me who is whose antecedent.}''

\subsubsection{Did participants access the source data for qualitative data analysis?}

Participants appreciated how the \textit{Details and Concepts Panel} and \textit{Node View} made it easy to locate quotable material (DG2-R4) when writing up their findings, as P1 shared: ``\textit{Most of the time we directly quote long sentences or paragraphs [...] for me the most convenient method is when I can see the original sentence in the sidebar when I select an indicator, which is quite good because when I write articles, I can directly use this sentence. I only pick individual words occasionally when writing discussion sections [...] but most of the time I need complete sentences in my document, so having sentences readily available is more effective.}''
I2 similarly expressed how the system facilitated the process of returning to source data to find supporting evidence: ``\textit{I will have certain categories and certain relationships in my mind [...] but then you still have to go back to the data to see if this proposed framework is supported [...] That's when I use the node view to find the specific quotes. This [system] will definitely help me access quotes more easily than Excel.}''

On the contrary, participants sometimes considered examining details secondary because they viewed causal discovery and hypothesis generation as the primary purpose of the system. 
For example, I5 only paid attention to details when outputs were unexpected: ``\textit{I don't know if there will be some researchers who might feel unsatisfied with something and want to see how it was originally done [...] I want to quickly know what the situation is, where the problem occurred [...] but that's probably mainly because I don't have any dissatisfaction with this story right now, so I don't feel like I have anything to investigate. I would only want to look at the original sentence if there's an edge that seems counterintuitive.}''

%% file: sections/6_discussion.tex
\section{Discussion}
\label{sec:discussion}

\subsection{Considerations and Implications of Designing Computational Tools for Qualitative Data Analysis}

\subsubsection{Computational tools improve analytical experience beyond efficiency gains}

Our findings resonate with prior literature establishing that computational tools can accelerate qualitative data analysis \cite{open_coding_visual_gao_2024, patat_gebreegziabher_2023, scholastic_hong_2022}, yet offer additional benefits beyond efficiency that contribute to the broader fields of computational social science and HCI.

\paragraph{\textcolor{darkred}{\textbf{Information ``redundancy'' reduces analytical anxiety.}}}

We found that when our system extracted and visually presented indicators, concepts, and/or causal relationships in the form of tables or network diagrams, some participants (e.g., I5, P4) felt more secure and experienced less \textit{fear of missing out} (i.e., a common anxiety associated with concerns about overlooking important data patterns), even when they regarded some of the extractions as excessive. 
This empirical finding suggests that comprehensiveness and coverage could be more valued than accuracy and precision by certain users (e.g., \textit{maximizers} in Herbert Simon's satisficing-maximizing framework \cite{maximizer_simon_1956}), \textbf{informing future work in two ways}: first, technical evaluations of computational tools that partially automate qualitative data analysis could incorporate metrics beyond accuracy alone, given that users do not universally prefer perfectly precise outputs; second, designers must still recognize the point at which information abundance transitions from supportive to overwhelming.

\paragraph{\textcolor{darkred}{\textbf{Cognitive scaffolding overcomes analytical inertia.}}}
Our feedback study revealed that some participants (e.g., F5, I4) found causal discovery and hypothesis generation within unstructured, chaotic textual data daunting and insurmountable, which could lead to the paralyzing blank-page problem (i.e., the mental barrier of starting from scratch) and avoidance behaviors; yet, our system that quickly extracts and visualizes causal relationships could help reduce this inertia and create more psychologically manageable processes of validation, adjustment, and enhancement. 
These insights suggest that certain users seeking fast and decent solutions (e.g., \textit{satisficers} in the satisficing-maximizing framework \cite{maximizer_simon_1956}) may benefit from computational tools that provide the \textit{activation energy} described in behavioral change theory \cite{activation_theory_prochaska_1983} and the \textit{momentum} effects found in priming research in cognitive psychology \cite{priming_tulving_1990} and Csikszentmihalyi's flow theory \cite{flow_csikszentmihalyi_1990}.
\textbf{Future systems could} include sliders or settings panels that allow users to customize the level of automation and visualization complexity.

\paragraph{\textcolor{darkred}{\textbf{Dual-level visualization enables dual-purpose use.}}}
Building on prior work that visualized causal relationships in both numerical \cite{causal_viz_2021, causal_viz_bae_2017, causal_viz_dang_2015, causal_viz_deng_2021} and qualitative data \cite{causal_system_powell_2025, causal_system_powell_2024, causal_system_husain_2021, causal_llm_jalali_2024, causal_llm_hosseinichimeh_2024}, our system used dual-level causal network diagrams to explore these relationships in qualitative data at both the conceptual level for an overview and theoretical insights, and the indicator level for technical details and evidence, serving both \textit{analytical} and \textit{communicative} purposes. 
In practice, as reported by feedback study participants (e.g., F5, P4, P10), this adaptability proves valuable across scenarios such as stakeholder presentations where intuitive understanding is more important than precision, academic talks that demand visual clarity, and publications requiring ready-to-use figures.
\textbf{Computational tool designers could further} consider integrating interactive community detection algorithms \cite{community_cruz_2014}, centrality measures \cite{centrality_freeman_1978}, and visualizations of uncertainty at both the indicator and concept level \cite{uncertainty_viz_pang_1997} into causal network interfaces to enhance analytical depth, rigor, and communicative clarity.

\paragraph{\textcolor{darkred}{\textbf{Interaction design creates companion-like experiences.}}}

Participants in our feedback study (e.g., F4, I4) found our system psychologically and emotionally supportive, as the process of reviewing and responding to extracted indicators and causal relationships, and collaboratively finishing mapping indicators to concepts and generating hypotheses, provided a sense of partnership. 
This finding echoes \textit{social presence theory} \cite{spt_short_1976}, which proposes that certain features like responsiveness (e.g., incorporating user-provided research overviews as context for indicator extraction), immediate feedback (e.g., storing user mappings of indicators to concepts within the concept list), and interactive exchange (e.g., iteratively re-visualizing causal networks) can create feelings of mutual presence, connection, and authentic co-existence that improve qualitative data analysis experiences. 
\textbf{It would be beneficial if future computational tools incorporate} anthropomorphic interface elements \cite{anthropomorphism_epley_2007} like conversational interactions to foster greater feelings of companionship rather than positioning users as operators of mechanical tools.

\subsubsection{Computational tools restructure analytical and cognitive processes}

Our feedback study echoes previous research showing that computational methods can facilitate causal discovery in qualitative data \cite{causal_llm_jalali_2024, causal_llm_hosseinichimeh_2024, causal_text_meng_2025}, while further revealing that these structured workflows do not always align with users' established practices.

\paragraph{\textcolor{darkred}{\textbf{From passive acceptance, confirmation to proactive cognitive forcing, exploration.}}}

Our findings revealed that some participants (e.g., P8, P9) reported a tendency to become lazy and less rigorous, conventionally analyzing qualitative data from source data to indicators, then to concepts, and finally to causal relationships; however, when using our system, they worked from indicators to concepts, then to causal relationships, only examining source data when outputs diverged from their expectations or theoretical knowledge. 
This analytical shift, combined with the system's presentation format that had users ``\textit{labeling}'' indicators via pre-structured tables, reinforced a ``\textit{task-like}'' mentality that led users to habitually move from one indicator to the next as if they were ``\textit{data labelers},'' potentially reframing qualitative data analysis as a process of validating system outputs rather than one of discovery, abstraction, and theorizing \cite{hci_qual_lazar_2017}.

Only a few participants (e.g., I5) were psychologically aware of their potential over-reliance on system outputs, and they deliberately cross-checked with source data. 
This self-imposed cognitive forcing \cite{cognitive_forcing_croskerry_2003} aligns with Kahneman's \textit{dual-process theory} \cite{theory_kahneman_2011}, wherein users consciously transition from \textit{System 1 thinking}, characterized by automatic acceptance of system outputs, to \textit{System 2 thinking}, which involves effortful, deliberate decisions.
\textbf{Therefore, future iterations of such computational tools could} encourage cognitive forcing behaviors by implementing adaptive, explainable causal discovery pipelines that embed epistemic transparency features, prompting users to examine how the system justifies causal relationships rather than passively accepting them.

\paragraph{\textcolor{darkred}{\textbf{From rigid, predetermined workflow to fluid, integrative thinking.}}}
We found that some feedback study participants (e.g., F3, P12) tend to categorize and causally connect qualitative data simultaneously, particularly when working with smaller datasets or familiar domains, engaging in a flexible analytical process. 
However, computational tools potentially ``\textit{domesticate}'' users by implementing predetermined workflows that dictate what the system executes first, what information users need to provide, and what outputs users receive. 
This could create an \textit{automation paradox} \cite{theory_bainbridge_1983, theory_parasuraman_1997}: users recognize the efficiency of computational tools, yet may choose not to use them because adapting to system workflows that do not perfectly align with their established practices and habituation can be cognitively demanding. 
\textbf{Moving forward, systems could allow users to} toggle between structured, step-by-step guidance and free-form analysis modes based on their level of expertise and the complexity of their data.

\paragraph{\textcolor{darkred}{\textbf{From sequential presentation to holistic processing.}}}
In our feedback study, some participants (e.g., I3, P9) used to skim through all the raw data first to form initial impressions and identify recurring patterns that crossed their attention thresholds \cite{theory_tversky_1974, theory_kahneman_2011}; however, the sequential, paginated presentation of our system (e.g., when displaying extracted indicators) led them to disproportionately allocate more time to the beginning and rush through the rest. 
This design could potentially serve as a form of algorithmic mediation \cite{theory_baumer_2024} that introduces selection bias in what gets flagged as indicators, framing bias in how these indicators are described, salience bias in which indicators receive emphasis, and sequence bias in the order in which indicators appear \cite{theory_tversky_1974}. 
More concerning is that, when the initial data happen to represent particular social groups, computational tools could inadvertently privilege certain voices or experiences over others. 
\textbf{An extension of our system would be to} incorporate randomization features or importance-based ordering instead of the default sequential ordering to mitigate these potential biases.

\subsubsection{Computational tools encounter paradigmatic boundaries in collaboration}

Our key findings align with previous work suggesting that computational tools should collaborate with rather than replace users in qualitative data analysis \cite{llm_inductive_lam_2024, llm_inductive_zenimoto_2024, llm_inductive_lee_2024, inductive_labelling_dai_2023, chalet_meng_2024}, and further reveal that the desired degree of autonomy varies considerably across research paradigms, particularly in interpretive and post-positivist traditions.

\paragraph{\textcolor{darkred}{\textbf{Balancing quantification and interpretation in interface design.}}}

Our findings suggest that some participants (e.g., F4, I4, P1, P12) placed less emphasis on the numerical visual elements (e.g., percentages of certain concepts) in our system.
This insight explicates the reductionist risks inherent in data visualization \cite{visual_salient_tufte_1983}, where user attention is automatically drawn to salient visual features (e.g., color, size, position) \cite{visual_salient_ware_2019}, and, more worryingly, as \textit{cognitive fit theory} \cite{cft_vessey_1991} indicates, visual design can alter users' mental representations of how to approach qualitative data analysis in ways that may conflict with interpretivist/critical paradigms that prioritize social justice and amplify marginalized voices.
These epistemological concerns \textbf{suggest that future computational tool designs could} provide narrative-centered displays and cues (e.g., prompts, annotations, and filtering options) to deliberately highlight outliers and minorities.

\paragraph{\textcolor{darkred}{\textbf{Negotiating analytical alignment with computational tools.}}}
We found that some participants (e.g., F4, P8) treated our system as a ``\textit{research assistant},'' expecting it to engage in exchanging and negotiating values in ways that resonate with \textit{common-ground theory} \cite{common_ground_clark_1991}, which posits that successful communication and collaboration require shared knowledge, beliefs, and assumptions.
Such needs are particularly pronounced in the interpretivist paradigm, where positionality is fundamental \cite{qda_tracy_2024}, as \textit{standpoint theory} \cite{standpoint_harding_1991} conceptualizes knowledge as situated, shaped by social positions that determine how the world is interpreted and what knowledge is accessible -- an aspect that computational tools may inherently lack.
\textbf{A promising direction for future research is therefore to} incorporate interactive onboarding processes that support mental model alignment \cite{mental_model_johnson_1983} between users and computational tools in qualitative data analysis.

\paragraph{\textcolor{darkred}{\textbf{Addressing paradigmatic concerns about causal inference.}}}

In our feedback study, some participants (e.g., P1, P9) trained in post-positivist methodologies questioned the use of qualitative data to examine causal relationships, contending that valid causal inference requires standard methods such as synthetic control \cite{sc_abadie_2010}, difference-in-differences \cite{did_angrist_2009}, structural models \cite{causality_pearl_2009, causal_spirtes_2000}, or randomized and quasi-experimental designs \cite{causal_shadish_2002}, and emphasizing that ``\textit{correlation does not imply causation}'' \cite{correlation_causation_bleske_2015}.
We acknowledge this concern and clarify that our system does not claim to identify definitive causation or establish robust theories based solely on qualitative data; rather, it helps discover putative, exploratory causal relationships to inform hypothesis generation and theory building, which should be further verified using methodologically appropriate approaches.

Besides, we note that exploring causal relationships in qualitative data could offer distinct advantages that complement quantitative causal inference, as such explorations preserve contextually rich, thick interpretation without being reductive.
For example, in MHStigmaInterview-20, one sentence stated ``\textit{I don't rent to those people because I don't trust strangers on my property. I worry about the upkeep. It's more about protecting my property value,}'' first presenting the outcome (i.e., refusing to rent), then discussing tangential concerns (i.e., distrust, housekeeping worries), and only later revealing the core reason (i.e., property devaluation); this layered discourse context helps uncover tacit causal relationships that are worth testing through quantitative methods.
Furthermore, computational tools that discover such causal structures have the potential to facilitate \textit{abductive reasoning} in qualitative inquiry, not just through partial automation, but also through co-exploration, where users and systems work together to discover and interpret causal relationships in discourse. 
\textbf{We remind designers of future computational tools for causal discovery in qualitative data to} attend carefully to their causal claims, recognize epistemological assumptions across paradigms, and acknowledge how such tools might transform interpretive research practices.

\subsection{Limitation and Future Research}

Our study has several limitations that point to opportunities for future research. 
First, the scope of our current evaluation could be extended.
Specifically, we grounded our qualitative feedback study in established practices for novel systems \cite{qualitative_eval_jo_2022, qualitative_eval_arawjo_2024, evaluate_olsen_2007}, yet the use of relatively short sessions and pre-selected datasets potentially limits our ability to capture how users would integrate such tools into their analytical practices over extended periods. 
Moreover, although we included a preliminary quantitative technical evaluation, we relied on pilot testing for prompt design rather than conducting systematic ablation studies. 
We also neither asked participants to explicitly document their hypotheses, nor validated hypothesis quality with experts in a controlled setting.
Future work should therefore bridge these gaps through 1) longitudinal studies using firsthand field data alongside case studies using open qualitative datasets to evaluate whether the system can help rediscover the findings; 2) rigorous ablations for prompt designs, particularly for indicator extraction and concept mapping; and 3) between-subjects experiments to systematically assess the quality of generated hypotheses.

Second, our system only focuses on within-sentence causal relationships, potentially missing those that span multiple sentences, omitting broader contextual patterns across discourse, and thus bounding the utility for theory building.
Advancing causal discovery and hypothesis generation in qualitative data analysis would greatly benefit from future research exploring systems that detect and visualize discourse-level causal relationships \cite{causal_multiple_wang_2024} and provide users the option to explore at either the sentence- or discourse-level.

And third, our feedback study highlighted the paradigm tension regarding semi-automated causal discovery as one of the primary concerns, an issue we currently touch upon only by initiating a conversation on how computational tools interact with interpretive inquiry.
Future research could thus develop a formal interaction model of computationally assisted theorizing, explicitly describing how system outputs reshape human interpretation.

Finally, we acknowledge the scalability limits when applied to massive corpora. 
Future iterations will need to address potential latency and visual clutter through optimization and aggregation strategies.

\section{Conclusion}

Through designing and evaluating \texttt{QualCausal}, this study demonstrates that computational tools could support users in analyzing qualitative data, specifically to explore causal relationships and generate hypotheses, by interactively constructing and visualizing causal networks. 
Our findings reveal that users value computational assistance that reduces their analytical workload, provides cognitive scaffolding, and preserves their interpretive autonomy and access to contextual source data. 
However, users also experience tension between efficiency and their research practices and paradigms. 
These results have broader implications for designing computational tools that align with the epistemologies and research paradigms across disciplines. 
We hope this study will contribute to the ongoing discussion about how computational tools can be thoughtfully integrated into qualitative data analysis.